\documentclass[english,aps,prb,amsmath,amssymb,superscriptaddress,twocolumn,a4paper]{revtex4-1}
\usepackage{amsmath}  \usepackage{amssymb}  \usepackage{amsthm}
\usepackage{graphicx} 
\usepackage{dcolumn}	
\usepackage[colorlinks=true,breaklinks]{hyperref}
\usepackage{longtable}
\usepackage{ulem}     
\usepackage{braket}

\begin{document}

\title{Non-local Coulomb interactions on the triangular lattice in the high-doping regime: Spectra and charge dynamics from Extended Dynamical Mean Field Theory }

\author{Sophie Chauvin}
\affiliation{Centre de Physique Th\'eorique, \'Ecole Polytechnique, CNRS UMR7644, Universit\'e Paris-Saclay, 91128 Palaiseau Cedex, France}
\affiliation{Laboratoire des Solides Irradi\'es, \'Ecole Polytechnique, CNRS,
CEA,  Universit\'e Paris-Saclay, F-91128 Palaiseau, France}
\author{Thomas Ayral}
\affiliation{ Physics and Astronomy Department, Rutgers University, Piscataway, NJ 08854, USA}
\affiliation{ Institut de Physique Th\'eorique (IPhT), CEA, CNRS, UMR 3681, 91191 Gif-sur-Yvette, France}
\author{Lucia Reining}
\affiliation{Laboratoire des Solides Irradi\'es, \'Ecole Polytechnique, CNRS,
CEA,  Universit\'e Paris-Saclay, F-91128 Palaiseau, France}
   \affiliation{European Theoretical Spectroscopy Facility (ETSF)}
\author{Silke Biermann}
\affiliation{Centre de Physique Th\'eorique, \'Ecole Polytechnique, CNRS UMR7644, Universit\'e Paris-Saclay, 91128 Palaiseau Cedex, France}
\affiliation{European Theoretical Spectroscopy Facility (ETSF)}
\affiliation{Coll\`{e}ge de France, 11 place Marcelin Berthelot, 75005 Paris, France}

\begin{abstract}
We explore the two-dimensional extended Hubbard model on the triangular lattice
in the high doping regime. On-site and nearest-neighbour repulsive interactions are treated in a non-perturbative way by means of Extended Dynamical Mean Field Theory. We compute the low-temperature phase diagram, displaying a metallic phase and a symmetry-broken phase for strong intersite repulsions. We describe the correlation effects on both single-particle and two-particle observables in the metallic phase. Whereas single-particle spectra feature a Hubbard satellite typical of strongly correlated systems, local susceptibilities remain close to their non-interacting limit, even for large on-site repulsions. We argue that this behaviour is typical of the strongly doped case. We also report a region in parameter space with negative static local screening.
\end{abstract}

\maketitle

Transition metal oxides display a wide variety of exotic phenomena as a function of the filling of the $d$-shell. Cuprates are the most popular example, going from a Mott-insulating phase to a charge-density wave 
to a superconductor to a metal \cite{imada}.
A similar phenomenology is realized for materials on the triangular lattice, such as sodium-doped cobaltates, Na$_x$CoO$_2$, where properties such as superconductivity~\cite{takada_nature_03} or a high thermoelectric power~\cite{terasaki_PRB_97} can be tuned with doping. An accurate description of materials away from the 
half-filled limit is thus important.

On the triangular lattice at incommensurate filling, charge dynamics can lead to a variety of exotic phases~\cite{tocchio_PRL_14}. Their intriguing properties originate from the interplay between local and non-local interaction parameters, as well as quantum hopping. Specifically, a phase exhibiting simultaneously charge-order and metallicity~\cite{tocchio_PRL_14} in the extended Hubbard model is reminiscent of the pinball liquid for spinless fermions~\cite{hotta_PRB_06, hotta_JPCM_07}, or the supersolid for hard-core bosons~\cite{wessel_PRL_05, heidarian_PRL_05, melko_PRL_05, zhang_PRB_11}, where the particles segregate between mobile and frozen. Understanding such phases is not only of purely academic interest, but also helps to describe actual materials more accurately. An anisotropic extended Hubbard model was used to describe charge-ordering in $\theta$-(BEDT-TTF)$_2$X~\cite{cano_PRB_11}. Multiorbital effects were shown to enrich again the phase diagram (relevant for materials with $e_g$ orbitals)~\cite{fevrier_PRB_15}.
The possibility of charge instabilities originating from the microscopic details of the hopping processes within a 3-orbital system was described in Ref.~\onlinecite{koshibae_PRL_03}.
Quite generally, it is expected that charge excitations and screening play a key role in the doped triangular lattice. 

The development of many-body techniques that are able to describe the physics of charge-ordering and screening, induced by long-range Coulomb interactions in strongly correlated electron materials, has become an active topic in condensed matter theory in recent years. Extended Dynamical Mean Field Theory (EDMFT) 
\cite{Sengupta1995,Kajueter1996,Si1996}
widens the scope of Dynamical Mean Field Theory (DMFT) 
(for a review, see Ref.~\onlinecite{GeorgesRMP96})
to systems with non-local interactions. In this method, the lattice model is mapped onto an effective local impurity problem {\it with frequency-dependent effective local interactions}, that are related self-consistently to the local polarization function. Following the development of improved Quantum Monte Carlo algorithms suitable for solving impurity models with frequency-dependent interactions, several works presenting fully self-consistent EDMFT implementations have appeared over the last few years (see e.g. Refs~\onlinecite{Ayral2012,Ayral2013, Huang2014, VanLoon2014, Hafermann2014, VanLoon2015, Ayral2017}). In a recent realistic study \cite{nomura-C60-review} this theory was used to determine the phase diagram of alcali-doped fullerides. Ref. \onlinecite{rademaker} investigated
glassy behavior, possibly relevant for organic materials.
Finally, Ref. \onlinecite{Camjayi2008} presented a DMFT study of the Wigner-Mott
transition on the triangular lattice.

Combined many-body perturbation theory + dynamical mean field theory (``$GW$+DMFT'') \cite{Biermann2003,Sun2004,Ayral2012,Ayral2013,Biermann2014} 
goes beyond the local description of screening by including the non-local polarization at the level of the random phase approximation.
The dual boson scheme \cite{Rubtsov2011,VanLoon2014,Stepanov2015} introduces non-local corrections beyond $GW$+DMFT, thereby curing artefacts of the plasmon dispersion present in the simpler schemes.
Most work so far has however focussed on systems on the square lattice,
and much less is known for triangular lattice models (see however Refs~\onlinecite{Hansmann2013,Hansmann2016}).

In this study, we explore single-particle properties and screening in an extended Hubbard model on the triangular lattice, focussing on the high-doping limit.
The paper is organized as follows: in section \ref{model}, we introduce the model and define our notations. Section \ref{methods} reviews briefly the main ideas of EDMFT and the scheme resulting from its combination with the Fock diagram. In section \ref{sec:phase_diag}, we present the phase diagram of the extended Hubbard model within these schemes, before analyzing one- and two-particle observables in sections \ref{section:one_part} and \ref{section:two_part} respectively. Finally, we present our conclusions.
In the Appendix, we rationalize the intriguing finding of a negative effective screened interaction, by discussing how this effect occurs in a simple model system.

\section{Model\label{model}}

\begin{figure}[htb]
\begin{center}
\includegraphics[width=1.0\columnwidth]{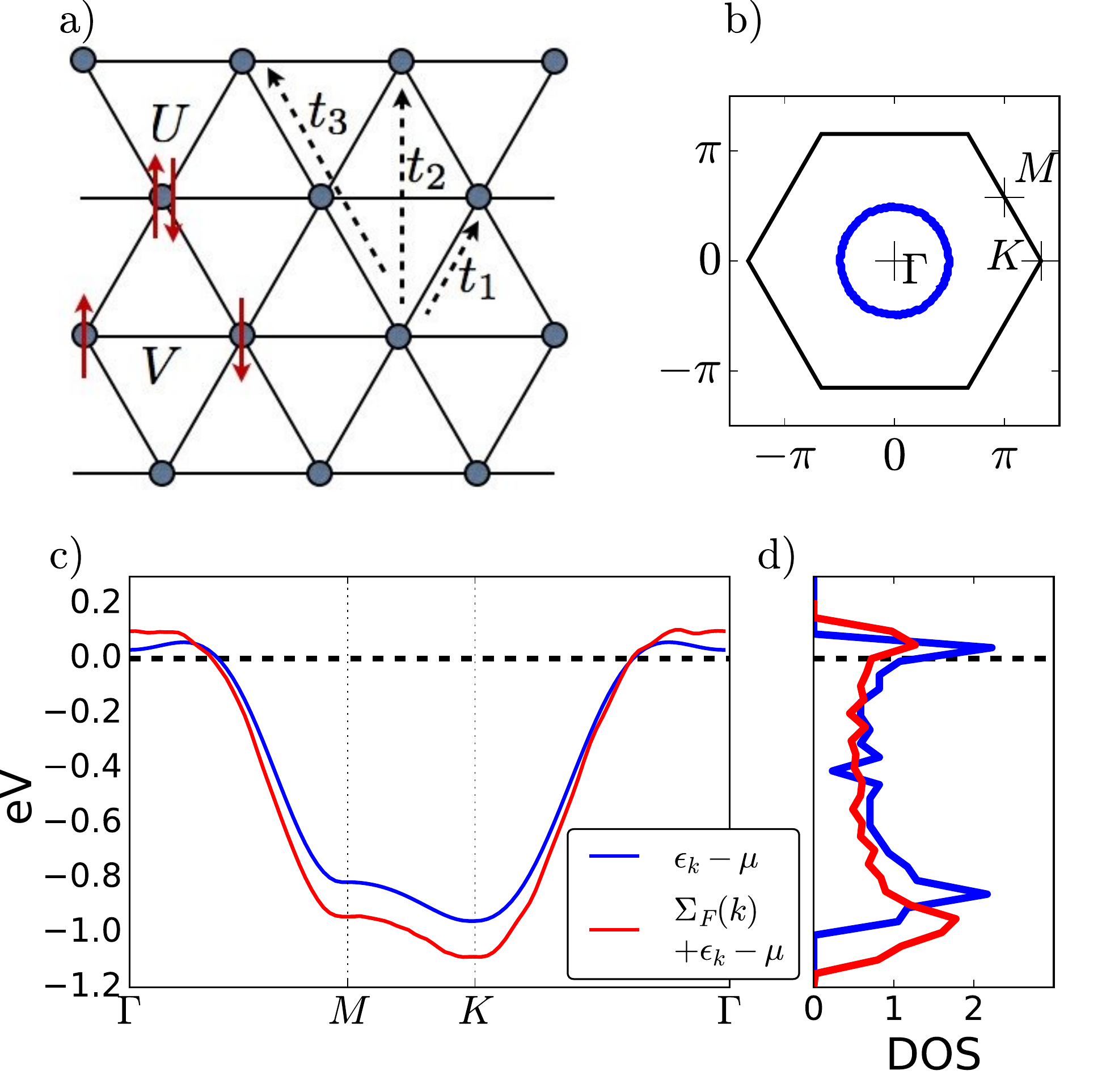}
\caption{ a) The triangular lattice, with the hopping and interaction parameters. The hopping is parametrized up to the third nearest neighbour and the interaction is parametrized up to the nearest neighbour. b) The 1st Brillouin zone of the triangular lattice, with high-symmetry points. Blue: the non-interacting Fermi surface. c) The bare dispersion $\varepsilon(\mathbf{k}-\mu)$, together with the dispersion modified by the Fock self-energy. The chemical potential lies close to the band edge, such that the filling per spin is $\braket{n_{\sigma}}=0.83$ (1.67 electrons per site). d) Corresponding densities of states. Note the two van Hove singularities, corresponding to a flattening in the band dispersion.}
\label{hoppings}
\end{center}
\end{figure}

The minimal model to study the interplay between local and non-local interactions, and the induced screening processes, is the extended Hubbard model, the Hamiltonian of which is given by:
\begin{equation}
\mathcal{H}  =
\sum_{i,j,\sigma} t_{ij} c_{i\sigma}^{\dagger} c_{j\sigma}
+ U \sum_i n_{i\uparrow} n_{i\downarrow}
+ V \sum_{\braket{i,j}} n_i n_j,
\label{hamiltonian}
\end{equation}
where $c_{i\sigma}^{\dagger}, c_{i\sigma}$ are electronic creation and annihilation operators, respectively, for site $i$ and spin $\sigma$; $n_i = n_{i\uparrow} + n_{i\downarrow} = c_{i\uparrow}^{\dagger} c_{i\uparrow} + c_{i\downarrow}^{\dagger} c_{i\downarrow}$ is the total density on site $i$. The sum $\sum_i$ runs over all lattice sites, while $\sum_{\braket{i,j}}$ runs over all nearest neighbour lattice bonds.
The first term of Eq.~\eqref{hamiltonian} is the non-interacting part and accounts for the hopping of the electrons between the sites. The second and third terms are the interacting part, and account for Coulomb repulsion. The second term is an on-site Hubbard repulsion term. The third term is a nearest-neighbour (non-local) Coulomb repulsion. Hence, the extended Hubbard model is able to capture the interplay between correlations induced by local and non-local interactions. 

The Hamiltonian~\eqref{hamiltonian} can be recast in the more compact form:
\begin{equation}
\mathcal{H}  =
\sum_{i,j,\sigma} t_{ij} c_{i\sigma}^{\dagger} c_{j\sigma}
+ \frac12 \sum_{i,j} v_{ij} n_{i} n_{j},
\label{hamiltonian_int}
\end{equation}
where we have defined the real-space interaction $v_{ij} = U \delta_{ij} + V \delta_{\braket{i,j}}$. The Fourier transform of $v_{ij}$ has the following expression:
\begin{equation}
v_{\mathbf k} = U + 2 V \left( \cos(k_x) + 2 \cos\left(\frac{k_x}2\right) \cos\left(\frac{\sqrt3 k_y}2\right) \right).
\label{fourier_interaction}
\end{equation}

Our model is depicted in Fig.~\ref{hoppings}(a), the corresponding
Brillouin zone in Fig.~\ref{hoppings}(b).
The non-interacting part of the Hamiltonian, Eq.~\eqref{hamiltonian}, is chosen such as to reproduce the band dispersion of a typical two-dimensional transition metal oxide, namely the LDA band structure of Na$_{2/3}$CoO$_2$, as presented by Piefke \emph{et al.}\cite{piefke_PRB_10}. In the Hamiltonian, Eq.~\eqref{hamiltonian}, the hopping parameters $t_{ij}$ are the inverse Fourier transform of $\varepsilon_{\mathbf k}$, represented in Fig.~\ref{hoppings}(c). More specifically, $\varepsilon_{\mathbf k}$ is parametrized up to the third nearest neighbour, with:
\begin{equation}
t_1 = -0.134 \text{ eV,  }\phantom{mn}
t_2 = +0.028 \text{ eV, }\phantom{mn}
t_3 = -0.024 \text{ eV},
\end{equation}
via the dispersion relation:
\begin{equation}
\begin{split}
\varepsilon_{\mathbf k} =&
2 t_1    \left(\cos(k_x)    +     2 \cos(k_x/2) \cos(k_y \sqrt3/2)\right)\\
     & + 2    t_2  \left(\cos(\sqrt3 k_y)+  2 \cos(3 k_x/2) \cos(k_y \sqrt3/2)\right)\\
     & + 2 t_3 \left(\cos(2k_x)+2\cos(k_x)\cos(\sqrt3 k_y)\right).
     \end{split}
\end{equation}
The hopping parameters $t_{ij}$ are in the range of 100 meV, but due to the high connectivity of the triangular lattice, the non-interacting bandwidth is approximately 1.1 eV, as can be seen in the non-interacting density of states in Fig.~\ref{hoppings}(d).

Since we aim at probing the system in the high-doping regime, we set 
the band filling to $\braket{n_{\sigma}}=0.83$ (corresponding to an average of 1.67 electrons per atomic site). Note that this particular filling corresponds to a large non-interacting density of states at the Fermi level.
The non-interacting Fermi surface is composed of a single hole pocket centered around the $\Gamma$ point.

In our study, we will vary the onsite interaction $U$ up to 4 eV and the intersite $V$ up to 0.7 eV. In the following, all energies are given in electron-volts.

\section{Methods \label{methods}}

\subsection{Computational schemes}

The extended Hubbard Hamiltonian, Eq.~\eqref{hamiltonian}, is a many-body Hamiltonian; physical observables can therefore only be computed by using approximations. Here, we take a strongly correlated perspective to study the interplay between local and non-local interactions, and quantum hopping. Also, we want to access single-particle properties (as given by the one-body Green's function and relevant for photoemission experiments) and screening (as encoded in the two-body Green's function) on the same footing. Whereas dynamical mean-field theory (DMFT) can describe the interplay between local interaction and quantum hopping, extended DMFT is able to account for non-local interactions. In the following, we review these two methods. Then we describe how to optionally include a non-local Fock diagram.

\subsubsection{Reminder on the (extended) dynamical mean-field approximation}

Dynamical mean-field theories aim at computing observables on the 
lattice through a self-consistent loop, using an Anderson impurity 
model as a reference system. In the following, the subscript ``imp" 
refers to impurity quantities, while the subscript ``loc" refers to 
the local (i.e., site-diagonal) part of lattice quantities.

Single-site dynamical mean-field theory (DMFT) assumes that the local 
part of the lattice Green's function can be 
generated as the solution of a local impurity problem embedded into 
a bath. The latter is determined self-consistently, using the impurity 
self-energy as an approximation to the full self-energy of the lattice 
problem in the calculation of the local lattice Green's function.

Via this construction, local quantum fluctuations on the single-particle 
level are treated to all orders in perturbation theory, resulting in a 
frequency-dependent self-energy~\cite{kotliar_PT_04}.
The self-energy $\Sigma_{\mathrm{imp}}$ and the impurity Green's function, 
$G_{\mathrm{imp}}$, are computed by solving a self-consistently determined 
Anderson impurity model:
\begin{equation}
\begin{split}
S^{\text{DMFT}}_{\text{imp}}[c^*, c] =& -\sum_{\sigma} \iint_0^{\beta} \mathrm{d}\tau \mathrm{d}\tau' c^*_{\sigma}(\tau) \mathcal{G}_0^{-1}(\tau-\tau') c_{\sigma}(\tau')\\
&+ \frac{U}2  \int_0^{\beta} \mathrm{d}\tau n(\tau) n(\tau).
\label{action_dmft}
\end{split}
\end{equation}
The path-integral action Eq.~\eqref{action_dmft} is expressed in imaginary 
time $\tau$ and $\beta$ is the inverse temperature. The Grassmann variables 
$c^*, c$ correspond to the fermionic operators $c^{\dagger}, c$ respectively.
We denote $n = \sum_{\sigma=\uparrow,\downarrow} c_{\sigma}^* c_{\sigma}$.

\begin{figure}[htb]
\begin{center}
\includegraphics[width=1.0\columnwidth]{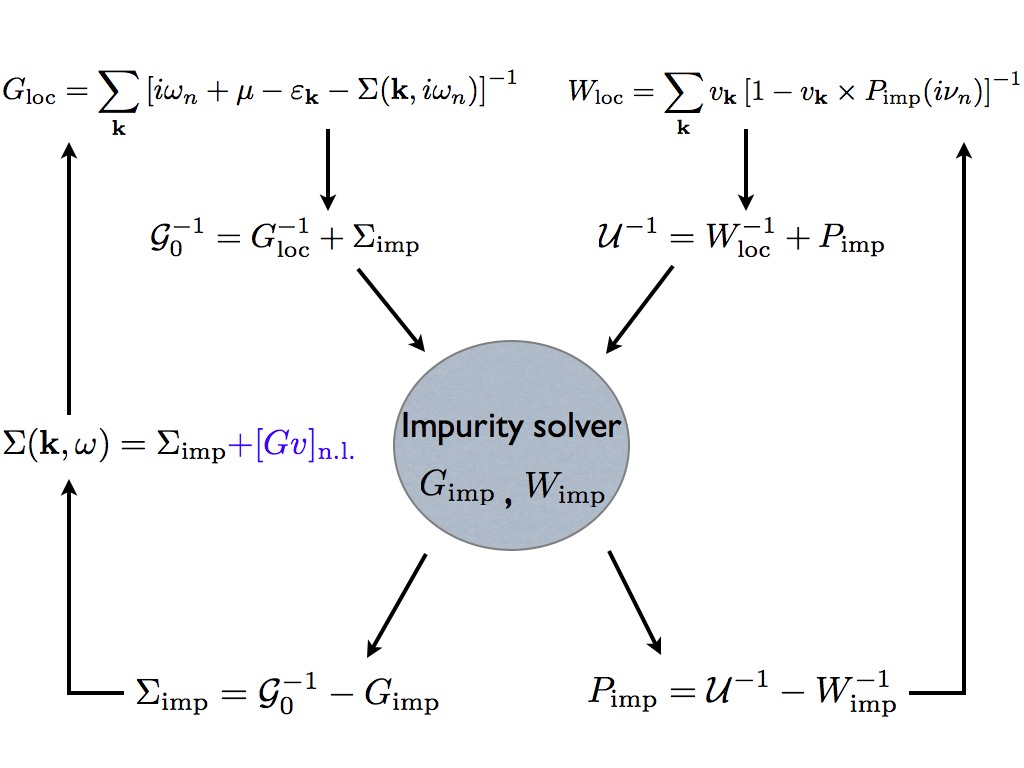}
\caption{The (Fock +) extended dynamical mean-field self-consistent loop. A non-local Fock term (blue) is optionally included in the self-energy, leading to the Fock+EDMFT scheme. Without the Fock term, the left part of the scheme 
reduces to the DMFT self-consistency loop.}
\label{edmft_loop}
\end{center}
\end{figure}

The dynamical mean field $\mathcal{G}_0$ is a self-consistent quantity. 
The self-consistency condition ensures that the local part of the lattice 
Green's function is given by the impurity Green's function, 
$G_{\mathrm{loc}} = G_{\mathrm{imp}}$, calculated under the approximation that 
the self-energy of the lattice is local and given by the impurity. The DMFT 
self-consistency loop is summarized on the left handside of 
Fig.~\ref{edmft_loop}.

DMFT becomes exact in the limit of infinite 
coordination~\cite{metzner_PRL_89, metzner_PRL_89_erratum}, where
non-local quantum fluctuations average out, resulting in a purely
local lattice self-energy. For a finite dimensional lattice, DMFT
yields an approximation that can be interpreted in two ways:
the straightforward interpretation consists in saying that the
lattice self-energy is approximated by a purely local quantity
generated from the impurity problem. This view can be expressed
by saying that the lattice self-energy has the following functional
dependence:
\begin{equation}
  \Sigma_{ij} = \delta_{ij} \,\Sigma_{\mathrm{imp}}
[G_{\mathrm{imp}}, U]
\end{equation}
where $i,j$ denote site indices. $i\omega$ (and later $i\nu$)  denote fermionic (resp. bosonic) Matsubara frequencies.  
A more subtle interpretation offers itself when restricting the
observables of interest to the local part of the lattice Green's 
function: DMFT assumes that this part can be generated from a
local impurity model, using the self-consistently calculated
bath. In the latter case, no explicit assumption on the lattice
self-energy is made, and the logic of the theory becomes comparable 
to the DFT construction, where a Kohn-Sham potential is calculated
for the mere purpose of generating the physical density, but without
any assumption on its relevance for the physical system.
From this perspective, the impurity model is the equivalent of the
Kohn-Sham system, and the construction of the local self-energy
from the impurity problem with the self-consistent bath the analogue
of the local density approximation to Kohn-Sham theory.

DMFT is designed to compute one-particle observables. Thus, 
just as Kohn-Sham theory yields the density but no other observables,
DMFT may not properly 
capture two-particle observables. Even more importantly, in the present
context, it disregards non-local interactions, and makes no attempt of
addressing screening processes induced by the latter.

In EDMFT, a bosonic propagator, $W$, is introduced, coupling to charge fluctuations introduced by a non-local interaction, $V$. Thus, EDMFT is designed to compute both one-particle and two-particle observables. Here, the two-particle observables are computed only in the charge channel. The bosonic propagator $W$ can be identified with the screened Coulomb interaction. It has a self-energy, $P$, the polarization (or irreducible polarisability). 

Again, the self-energy, impurity Green's function, the polarization $P_{\mathrm{imp}}$ and the screened interaction of the impurity $W_{\mathrm{imp}}$, are computed self-consistently by solving a local model, now featuring an effective bare local interaction, $\mathcal{U}(i\nu_n)$:
\begin{equation}
\begin{split}
S^{\text{EDMFT}}_{\text{imp}}[c^*, c] = &-\sum_{\sigma} \iint_0^{\beta} \mathrm{d}\tau \mathrm{d}\tau' c^*_{\sigma}(\tau) \mathcal{G}_0^{-1}(\tau-\tau') c_{\sigma}(\tau')\\
& + \frac{1}{2}
\iint_0^{\beta} \mathrm{d}\tau \mathrm{d}\tau' \mathcal{U}(\tau-\tau') n(\tau) n(\tau').
\end{split}
\label{action_EDMFT}
\end{equation}
EDMFT treats non-local interactions, $v_{ij}$, by transforming them into a frequency-dependent local interaction, $\mathcal{U}(i\nu_n)$~\cite{gatti_PRL_07}. $\mathcal{U}$ and $\mathcal{G}_0$ are both determined self-consistently. The self-consistency condition is now that both local quantities, $G_{\mathrm{loc}}$ and $W_{\mathrm{loc}}$, are given by the impurity: $G_{\mathrm{loc}} = G_\mathrm{imp}$ and $W_{\mathrm{loc}} = W_{\mathrm{imp}}$, under the approximation that the self-energy and the polarization are local and given by the impurity. The EDMFT self-consistency loop is summarized in Fig.~\ref{edmft_loop}.

As DMFT, EDMFT is a non-perturbative scheme, particularly well suited
for systems with strong interactions, dominated by local
self-energy and polarization processes. 
Similarly to the first interpretation of DMFT (see above), single-site
EDMFT can be 
interpreted as an approximation that assumes both the fermionic self-energy 
$\Sigma$ and the polarization $P$ to be local and to be given by the sum 
of all local diagrams. This can be expressed by the functional dependence 
of the EDMFT self-energy and polarization:
\begin{subequations}
\begin{align}
&\Sigma_{ij}(i\omega) = \delta_{ij} \,\Sigma_{\mathrm{imp}}
[G_{\mathrm{imp}}, W_{\mathrm{imp}}](i\omega), \\
&P_{ij}(i\omega) = \delta_{ij} \, P_{\mathrm{imp}}
[G_{\mathrm{imp}}, W_{\mathrm{imp}}](i\omega).
\end{align}
\end{subequations}
When adopting this interpretation, not only local quantities
(local Green's function and local screened interaction) become
accessible to calculations but also their full lattice counterparts
can be deduced from the self-consistent self-enery and polarization.

Analogously to the discussion above, one can also define a
  ``purist's point of view'', where only the arguments of the EDMFT
  approximation to Almbladh's $\Psi[G,W]$
  functional~\cite{almbladh}, namely the local Green's function $G_{loc}$ and the
  local screened Coulomb interaction $W_{loc}$, are accessible
  quantities, and the local self-energy $\Sigma_{loc}$
  and the local polarization $P_{loc}$ are considered as
  fictitious functions at the same level as the Kohn-Sham potential
  of density functional theory. Indeed, mathematically speaking,
  $\Sigma_{loc}$ and $P_{loc}$ enter the theory as Lagrange multipliers
  imposing physical values to $G_{loc}$ and $W_{loc}$.
  This standpoint highlights that, strictly speaking,
  deducing lattice Green's functions or lattice susceptibilities
  goes beyond the EDMFT framework in the same way as interpreting
  Kohn-Sham band structures of DFT as excitations of a system lies
  outside the initial purpose of DFT.

Here, we will adopt a hybrid point of view: we will indeed calculate
and analyze the lattice screened interaction and related quantities,
though we will not make use of lattice Green's functions.
 One may consider this approach as
analogous to the current practice of analyzing Kohn-Sham band structures 
from DFT, or momentum-resolved spectral functions from DMFT (even
though it is likely that quite generically the approximation of a 
local polarization is more severe than the one of a local self-energy).
Especially in two dimensions, where nonlocal effects beyond mean field
are expected to be quite large, the local approximation to the polarization
function is certainly a strong assumption.
EDMFT should therefore
be regarded as an exploratory tool to get qualitative insights 
into the charge dynamics. 

For completeness, let us mention that several extensions of EDMFT
to describe nonlocal self-energy and polarization processes have been
devised in recent years. The $GW$+EDMFT method \cite{Biermann2003,Sun2004}
supplements the local self-energy and polarization diagrams with the
nonlocal contributions from the $GW$ self-energy and $GG$ bubble
polarization. The former contains non-local screening processes and
the latter captures, in particular, nesting features of the Fermi
surface. The first self-consistent implementation of the $GW$+EDMFT
method has been developed in the context of square and cubic lattices
at half-filling \cite{Ayral2012,Ayral2013}. Recently, the nonlocal
Fock correction to the self-energy (which is also contained in GW)
has been shown to yield important
corrections to the phase boundary to the charge-ordered phase for
large values of the local interaction \cite{Ayral2017} (which motivates
the approximation we introduce in the next subsection). Furthermore,
thanks to its relative simplicity, the $GW$+EDMFT method has also
been applied to a number of realistic materials, ranging from systems
of adatoms on semiconducting surfaces \cite{Hansmann2013,Hansmann2016}
to the prototypical example of correlated materials, $\mathrm{SrVO}_{3}$
\cite{TomczakEPL100-12, TomczakPRB90-14, Boehnke2016}. 

Further extensions like the dual boson method \cite{Rubtsov2011,VanLoon2014,Stepanov2015}
have been proposed to remedy some of the shortcomings of EDMFT and
$GW$+EDMFT, like a poor description of collective modes \cite{Hafermann2014}
and inconsistencies between different ways of computing susceptibilities
\cite{VanLoon2015}. These approaches, in their full-fledged form,
require the (costly) computation of impurity vertices with three and
four external legs, making them 
hard to apply
in realistic settings
or
to map out entire phase diagrams with state-of-the-art algorithms.
Therefore, simplified versions
of these methods are emerging \cite{Stepanov2016} which do not require
the computation of these vertices. Interestingly, they yield phase
diagrams very similar to those obtained in $GW$+EDMFT \cite{Ayral2017},
and qualitatively at least, in EDMFT.

\subsubsection{Combining extended dynamical mean-field theory with
a non-local Fock diagram}

On top of the purely local self-energy diagrams treated by DMFT or EDMFT, we 
explore the addition of a non-local Fock term. Indeed, let us have a look at the first perturbative self-energy diagrams of the extended Hubbard Hamiltonian, Eq.~\eqref{hamiltonian}. The Hartree and Fock self-energies are, respectively:
\begin{align}
&\Sigma^H_{ij, \sigma}(i\omega) = \Sigma^H_{ii, \sigma} \delta_{ij} =  \left(U n_{i, \bar{\sigma}} + V \sum_{\braket{i,l}} n_l\right) \delta_{ij},\\
&\Sigma^F_{ij, \sigma}(i\omega) = - \delta_{\braket{ij}} V G_{ij, \sigma}(\tau=0^+),
\end{align}
The Hartree terms are local and static, so they can be absorbed in a redefinition of the chemical potential, $\tilde{\mu} = \mu - \frac{U}2 \sum_{\sigma} n_{i \sigma} - V \sum_{\braket{i,l}, \sigma} n_{l \sigma}$, provided that the charge density is homogeneous. The Fock term, however, is static but non-local and, by definition, is not included in the DMFT or EDMFT schemes.

The Fock term contributes a static, non-local, real part to the self-energy. Thus, at self-consistency, it can be viewed as a modification of the bare dispersion, $\varepsilon_{\mathbf k}$. In practice, it leads to a widening of the non-interacting band, Fig.~\ref{hoppings}. To this renormalized effective non-interacting band, it is possible to ascribe a density of states. At the Fermi level, the renormalized density of states is lower than the bare density of states.

Hence, the usual EDMFT scheme does not include all the first-order diagrams in $V$. In the Fock+EDMFT scheme, we make sure that at least the Hartree-Fock diagrams are present, on top of EDMFT. Fock+EDMFT can be viewed as a poor man's $GW$+EDMFT. Concerning the description of screening it goes beyond Hartree-Fock, because some screening is included via local polarization processes. 
We also note the relation to the recent ``Screened Exchange Dynamical Mean
Field Theory'' (``SEx+DMFT'') \cite{vanRoekeghem-PRL2014, vanRoekeghem-EPL2014, vanRoekeghem-CaFe2As2-PRB2016}, an approximation to $GW$+DMFT, where the non-local $GW$ self-energy is replaced by a screened exchange term.

\subsection{Implementation details}
The algorithm of Fig.~\ref{edmft_loop} is implemented using the TRIQS toolbox \cite{Parcollet2014}.
The impurity problem (Eq.~\eqref{action_EDMFT}) is solved with a continuous-time quantum Monte-Carlo solver (CTQMC \cite{Rubtsov2011}), using a hybridization expansion \cite{Werner2006} in the segment picture, with frequency-dependent interactions $\mathcal{U}(i\nu_n)$ \cite{Werner2009}.

The impurity solver computes the fermionic self-energy, $\Sigma_\mathrm{imp}(i\omega)$ using improved estimators\cite{Hafermann2012} and the polarization from the charge susceptibility:
\begin{equation}
\chi_{\mathrm{imp}}(\tau) = \braket{\mathcal{T}_{\tau} n(\tau) n(0)} - \braket{n}^2
\end{equation}
using the expression:
\begin{equation}
P_{\mathrm{imp}}(i\nu_n) = -\frac{\chi_{\mathrm{imp}}(i\nu_n)}{1 - \mathcal{U}(i\nu_n) \chi_{\mathrm{imp}}(i\nu_n)}.
\end{equation}

All calculations are made assuming a paramagnetic solution, i.e. by symmetrizing the up- and down-spin components in the self-energy. The calculations use a single-site impurity and assume a homogeneous solution.
The first Brillouin zone is discretized with $32 \times 32$ k-points.
In the calculations including the Fock term, a mixing of
50\% on the local self-energy is used to stabilize the iterative cycle and finally achieve convergence. 
Analytic continuations are carried out using using the Maximum Entropy algorithm in Bryan's implementation\cite{Bryan1990}.

\section{Phase diagram}
\label{sec:phase_diag}

\begin{figure}[htb]
\begin{center}
\includegraphics[width=1.0\columnwidth]{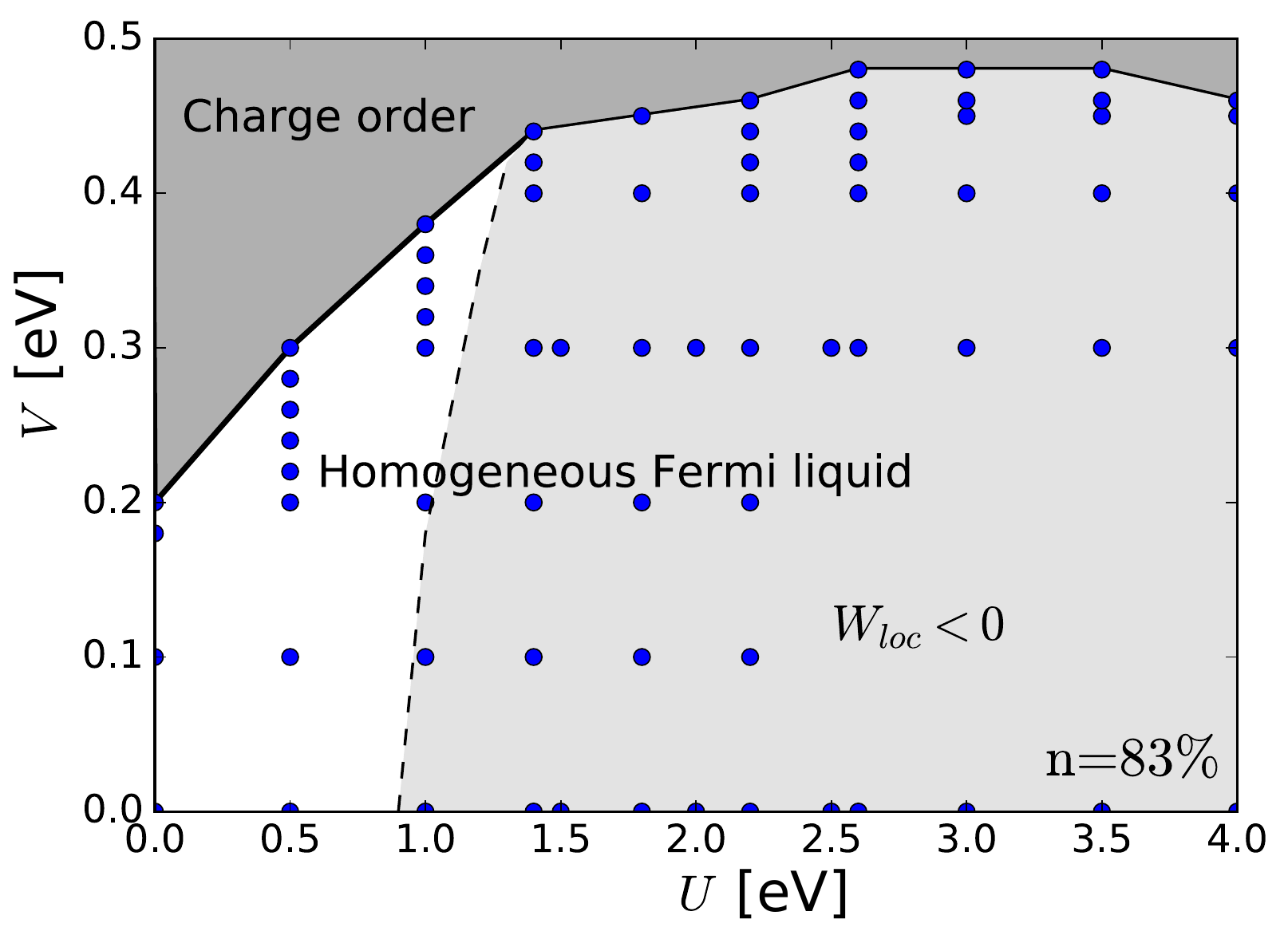}
\caption{The EDMFT phase diagram of the extended Hubbard model on a triangular lattice, at filling per spin $\braket{n_{\sigma}}=0.83$ and inverse temperature $\beta=100$ for the paramagnetic case. It displays a second-order phase transition, between a homogeneous Fermi-liquid phase (at low $V$) and a charge-ordered region (at high $V$, dark grey). The homogeneous Fermi-liquid phase is divided into a positive static screening region (at low $U$, white) and a negative static screening region (at high $U$, light grey).
The dots indicate the parameters used in the various runs that we performed.}
\label{phase_diag}
\end{center}
\end{figure}

\begin{figure}[htb]
\begin{center}
\includegraphics[width=1.0\columnwidth]{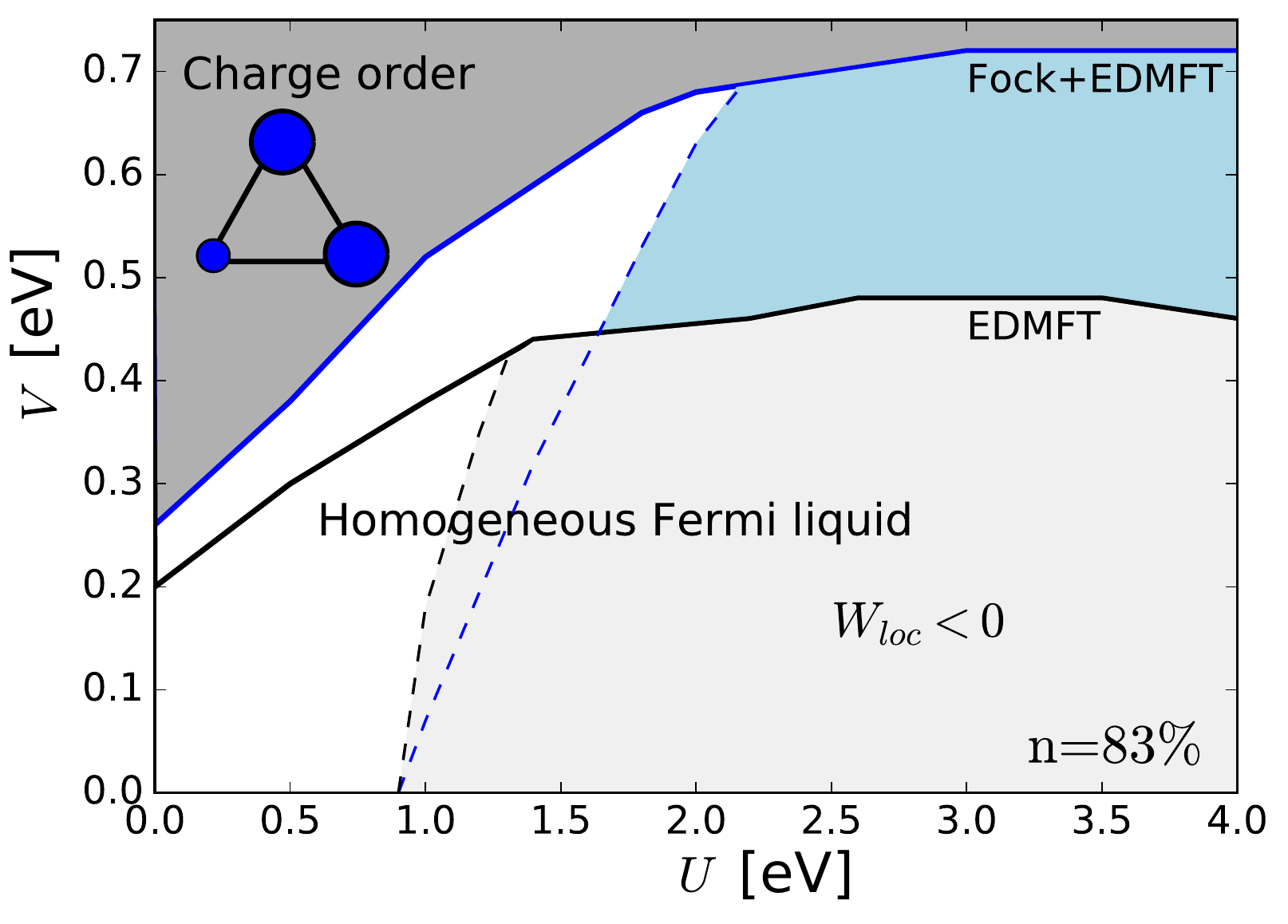}
\caption{Comparison between EDMFT and Fock+EDMFT phase diagrams of the extended Hubbard model on a triangular lattice, at filling per spin $\braket{n_{\sigma}}=0.83$ and inverse temperature $\beta=100$ (homogeneous paramagnetic calculations). It displays a phase transition, between a homogeneous Fermi-liquid phase (at low $V$) and a charge-ordered phase (at high $V$, dark grey). The homogeneous Fermi-liquid phase is divided into a positive static screening region (at low $U$, white) and a negative static screening region (at high $U$, light grey/blue). The inset shows the charge-ordering pattern, where two atoms retain two electrons each and one atom retains one electron.}
\label{phase_diag_comp_fock}
\end{center}
\end{figure}

In this section, we present the phase diagram (see Figs.~\ref{phase_diag} and \ref{phase_diag_comp_fock}) of the triangular lattice at filling per spin $\braket{n_{\sigma}}=0.83$ and inverse temperature $\beta=100$ (corresponding to a temperature of 116 K), computed within (Fock +) EDMFT, in the homogeneous paramagnetic case. 
This phase diagram displays two regions: at low $V$, a region where a homogeneous
metallic solution is found and a region where such a solution is no longer
stable, giving way to one or possibly more charge-ordered phases.

\subsection{Homogeneous metallic region}

At small intersite interaction $V$, the system is in a homogeneous Fermi-liquid phase. 
The metallic character extends to high Hubbard interactions $U$, because the system is strongly doped (the single band being filled at 83\%). 

The Fermi-liquid phase is divided into two sub-regions, depending on the value of the computed local static screening, $W_{\mathrm{loc}} (\omega = 0)$ (more details are provided in Sec.~\ref{section:two_part}). For low enough values of $U$, the local static screening is found to be positive: $W_{\mathrm{loc}} (\omega = 0) > 0$. However, for higher values of $U$, the local static screening is found to be negative: $W_{\mathrm{loc}} (\omega = 0) < 0$ (see shaded region in Fig.~\ref{phase_diag_comp_fock}).

\subsection{Transition line}

\begin{figure}[!htb]
\begin{center}
\includegraphics[width=1.\columnwidth]{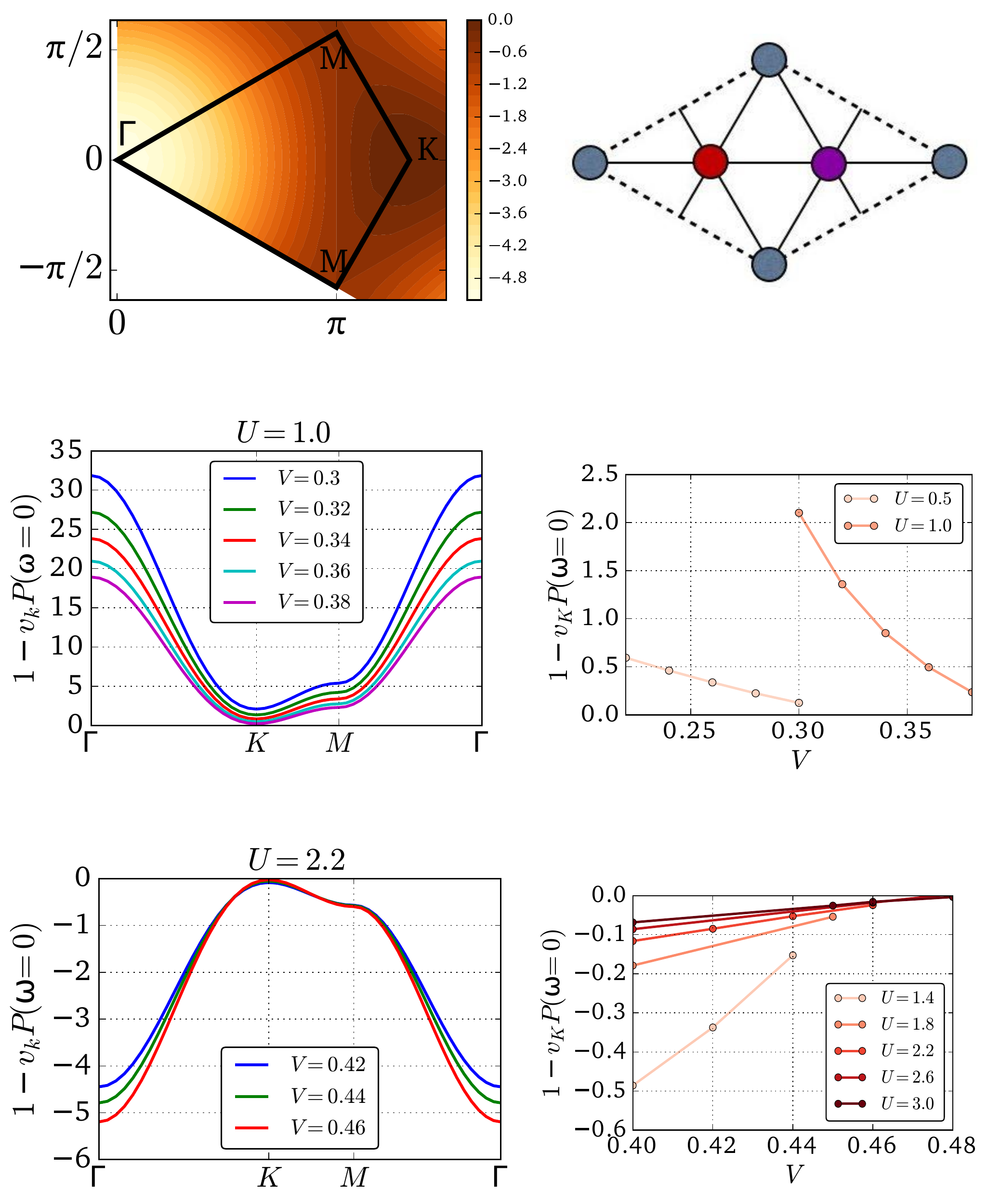}
\caption{a) The static value of $1 - v_{\mathbf q} P_{\mathrm{imp}}(\omega=0)$ in the irreducible Brillouin zone of the triangular lattice, for $U=2.2$ eV, $V=0.46$ eV, $\beta=100$ eV$^{-1}$, computed within EDMFT. The high-symmetry points $\Gamma$, $K$ and $M$ are represented. The $\mathbf{q}$-point where $1 - v_{\mathbf q} P_{\mathrm{imp}}(\omega=0)$ goes to 0 is the point where the divergence of the static susceptibility, $\chi(\mathbf{q}, \omega=0)$, sets in. b) and d) The static value of the dielectric function $1-v_{\mathbf q}P_{\text{imp}}$ on a $\mathbf{q}$-point path. Note the sign change of the dielectric function with respect to $U$. The positive sign occurs on the low-$U$ region of the phase diagram (see white region in Fig.~\ref{phase_diag}); the negative sign occurs in the high-$U$ region (see light-grey region in Fig.~\ref{phase_diag}). c) and e) The value of the static dielectric function for the ordering vector $\mathbf{q}_{\text{CO}}=K$, for various values of $U$.}
\label{co_Kpoint}
\end{center}
\end{figure}

The Fermi-liquid phase is separated from a charge-ordered region, for high values of the intersite interaction $V$. In this region of parameter space, calculations within homogeneous (Fock +) EDMFT could not be converged. The separation line in Fig.~\ref{phase_diag_comp_fock} marks the last converged point. More specifically, the separation line between the two phases can be monitored by the divergence of the static susceptibility $\chi(\mathbf{q}_{\text{CO}}, \omega=0)$, at some ordering vector $\mathbf{q}_{\text{CO}}$. The lattice susceptibility is linked to the bosonic propagator on the lattice: $W(\mathbf{q}, i\omega) = v_{\mathbf q} - v_{\mathbf q} \chi(\mathbf{q}, i\omega) W(\mathbf{q}, i\omega)$. 
The EDMFT lattice susceptibility is given by:
\begin{equation}
\chi(\mathbf{q}, i\omega) = - \frac{P_{\mathrm{imp}}(i\omega)}{1 - v_{\mathbf q} P_{\mathrm{imp}}(i\omega)},
\label{susceptibility_edmft}
\end{equation}
where the $\mathbf{q}$-dependence on the right-hand side is only present in the Fourier-transform of the interaction, $v_{\mathbf q}$, as given by Eq.~\eqref{fourier_interaction}. 
The divergence of the static susceptibility $\chi(\mathbf{q}, \omega=0)$ always occurs at $v_{q_{CO}}= 1/P_\mathrm{imp}(0)$ which leads for our model to an 
ordering vector $\mathbf{q}_{\text{CO}} = K$ (see Fig.~\ref{hoppings} and Fig.~\ref{co_Kpoint}).

The divergence condition $1 - v_{\mathbf q} P_{\mathrm{imp}}(i\omega) = 0$
for the susceptibility (Eq.~\eqref{susceptibility_edmft}) is equivalent to 
the fact that
the static dielectric function goes to 0, suggesting an instability to occur.
The static dielectric function is monitored in Fig.~\ref{co_Kpoint} in the vicinity of the charge-ordering transition. 

On top of going to zero at the phase transition, the static dielectric function changes sign depending on the region of the phase diagram (see shaded region in Fig.~\ref{phase_diag_comp_fock}, denoting negative static local screening).

\subsection{Charge-ordered region}

The ordered phase(s), at larger $V$, cannot be accessed via homogeneous single-site EDMFT, because of the symmetry breaking. At the specific commensurate doping $0.833 = \frac12 + \frac13$, the filling is suggestive of a $\sqrt3 \times \sqrt3$ ordering, with 3 atoms per supercell: 2 atoms are completely filled and one atom retains one electron (see Fig.~\ref{phase_diag_comp_fock}).

\subsection{Comparison of the EDMFT and Fock+EDMFT phase diagrams}

Upon addition of the Fock term, the charge-order transition line is pushed up in $V$. This is linked to a decrease in $\chi(\omega=0)$, connected to a decrease in the renormalized density of states at the Fermi level, $N(\epsilon_F)$, as will be explained in section~\ref{section:two_part}.

In the remaining part of this study, we analyze the homogeneous Fermi-liquid phase. We give a description of correlation effects by analyzing single-particle observables in section~\ref{section:one_part} and two-particle observables in section~\ref{section:two_part}.

\section{One-particle observables}
\label{section:one_part}

In this section, we describe the single-particle observables within EDMFT, in the Fermi-liquid phase. We argue that they are typical of a (moderately) correlated system, as they feature a lower Hubbard band.

\begin{figure}[htb]
\begin{center}
\includegraphics[width=1.0\columnwidth]{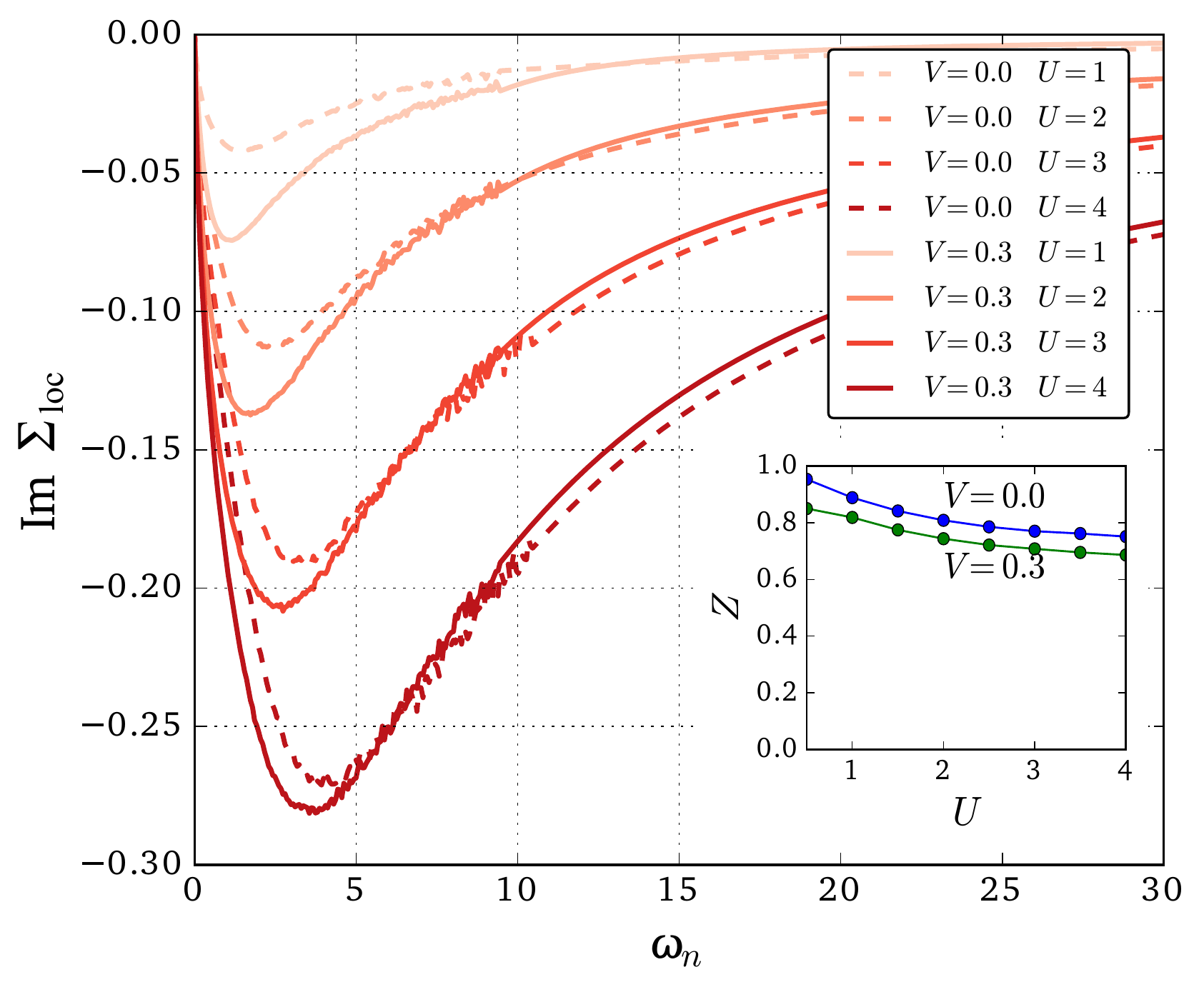}
\caption{The imaginary part of the impurity self-energy on the Matsubara axis, Im $\Sigma_{\text{imp}}(i\omega_n)$ within EDMFT, for various values of $U$ and $V$, and the corresponding quasi-particle renormalization factor, $Z$. The numerical values are typically those of a moderately correlated system. Correlations increase when either $U$, the on-site interaction, or $V$, the nearest-neighbour interaction, increases.}
\label{sigma}
\end{center}
\end{figure}
The self-energies on the Matsubara axis are depicted in Fig.~\ref{sigma}, for various values of $U$ and $V$. The self-energies are metallic and Fermi-liquid-like (since the imaginary part of the self-energy goes linearly to zero at $\omega=0$), even for high values of the Hubbard interaction $U$. A Mott metal-insulator transition is thus prevented by the strong doping, as expected. Correlations increase when either $U$ or $V$ increases, as can be seen from the quasi-particle renormalization factors in Fig.~\ref{sigma}. The quasi-particle renormalization factor $Z$ is defined as:
\begin{equation}
Z = \frac1{1- \left.\frac{\partial\Sigma(i\omega)}{\partial (i\omega)}\right|_{\omega=0}}.
\end{equation}
The values of the renormalization factor are typical of (weakly) correlated regimes, even at strong local interaction $U$. It is worth noting that the correlations increase when the nearest-neighbour interaction $V$ increases. This trend is opposite to what was observed for the half-filled extended Hubbard model on a square lattice \cite{Ayral2013}, where correlations decrease when $V$ increases. In Ref.~\onlinecite{Ayral2013}, it was argued that the intersite interaction $V$ effectively reduces the on-site interaction $U$ to the screened value $U_{\text{eff}} = \mathcal{U}(i\nu_n=0)$. In our case, the effect of $V$ is to enhance correlations, probably because
of the energetic cost associated to hopping processes taking place on sites
that are nearest neighbours to occupied sites.

\begin{figure}[htb]
\begin{center}
\includegraphics[width=1.0\columnwidth]{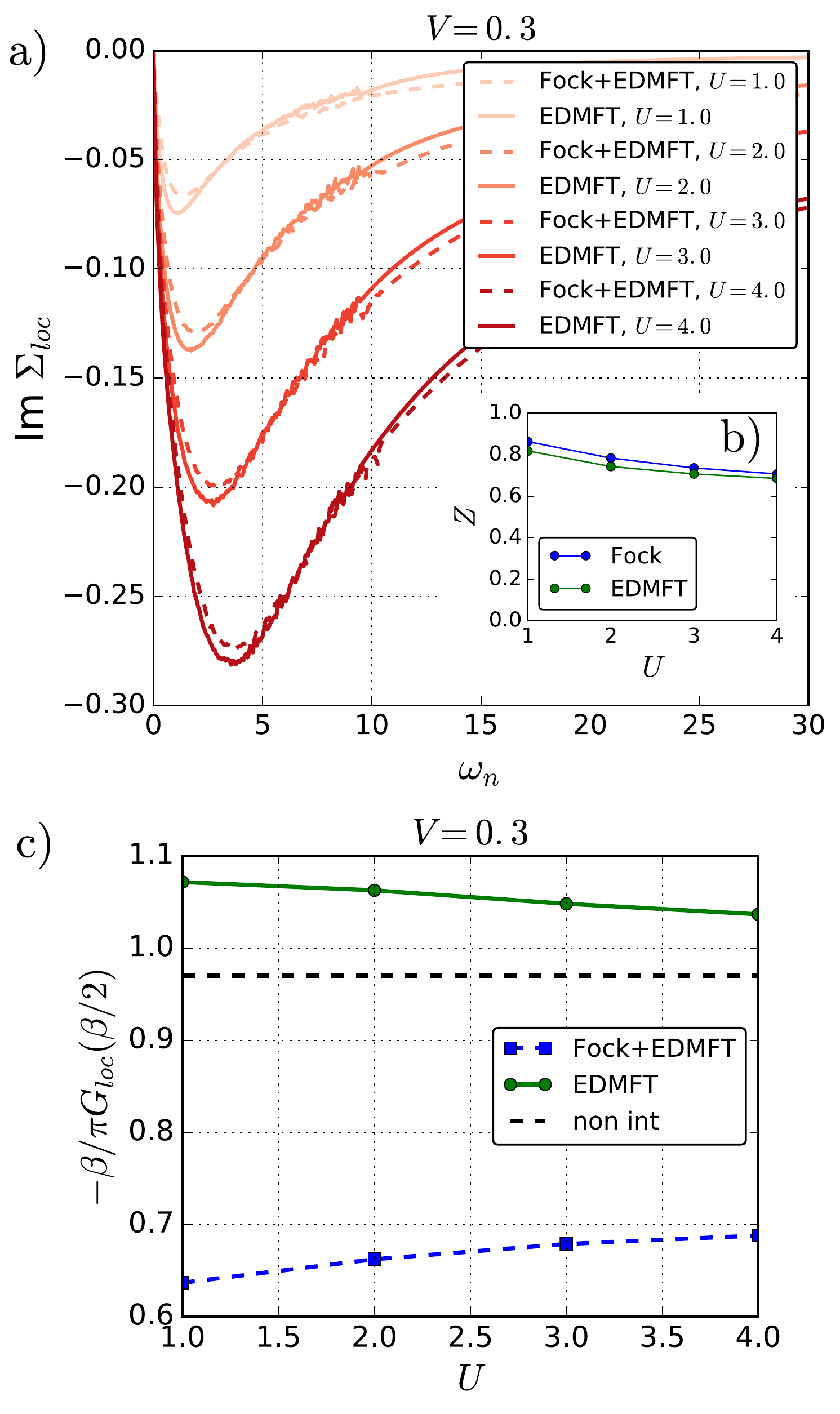}
\caption{Comparison between EDMFT and Fock+EDMFT for various one-particle quantities. a) Imaginary part of the impurity self-energy on the Matsubara axis. b) Corresponding renormalization factors. c) Density of states at the Fermi level, computed as $N(\varepsilon_F) \approx-\frac{\beta}{\pi} G_{\text{loc}}(\frac{\beta}{2})$. One of the effects of the Fock term is to reduce the interacting density of states at the Fermi level, as compared to both EDMFT and the non-interacting system.}
\label{single_part_fock}
\end{center}
\end{figure}
Fig.~\ref{single_part_fock} shows a comparison between EDMFT and Fock+EDMFT for some single-particle quantities. Concerning self-energies and renormalization factors, adding a Fock non-local term does not change the result significantly. The densities of states at the Fermi level, estimated as $N(\epsilon_F) \approx -\beta/\pi G_{\text{loc}}(\beta/2)$, are, however, different. This is an effect of the reduction of the effective density of states when going from EDMFT to Fock+EDMFT (see Fig.~\ref{hoppings} and the widening of the effective non-interacting band).

\begin{figure}[htb]
\begin{center}
\includegraphics[width=1.0\columnwidth]{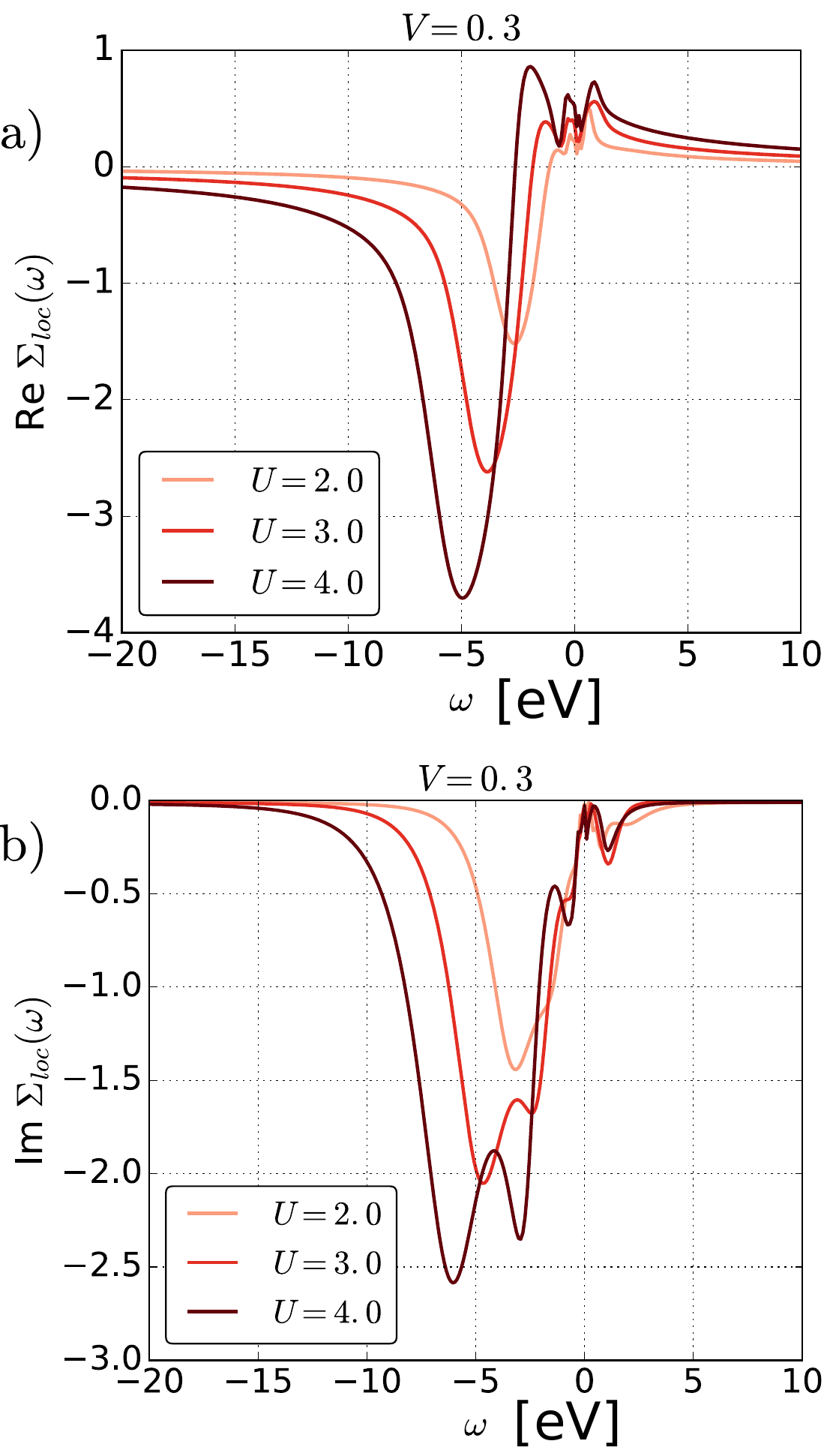}
\caption{Self-energy as a function of real frequency, obtained
from analytic continuation of the imaginary frequency data:
(a) real and (b) imaginary parts of the self-energy for $V$=0.3 and
varying $U$.}
\label{continuation_sigma}
\end{center}
\end{figure}

In Fig.~\ref{continuation_sigma}, we display the self-energies
on the real frequency axis, as obtained from the Matsubara frequency
data via an analytic continuation using the maximum-entropy algorithm.
The imaginary part is negative, as required by causality, and its absolute
value represents (up to a factor of $\pi$) the inverse lifetime of 
excitations. It takes on small values around the Fermi level (the
origin of the frequency axis), corresponding to the Fermi liquid nature of
the metallic phase investigated here. Its most prominent property
is the pronounced asymmetry of filled and empty parts of the spectrum,
corresponding to the large doping, inducing much shorter lifetimes for
hole excitations than for electrons. Interesting to note is also the
frequency scale on which the self-energy varies, namely between -10
and 2 eV, corresponding to the energy scale of the spectral function
(see below).

\begin{figure}[t]
\begin{center}
\includegraphics[width=1.0\columnwidth]{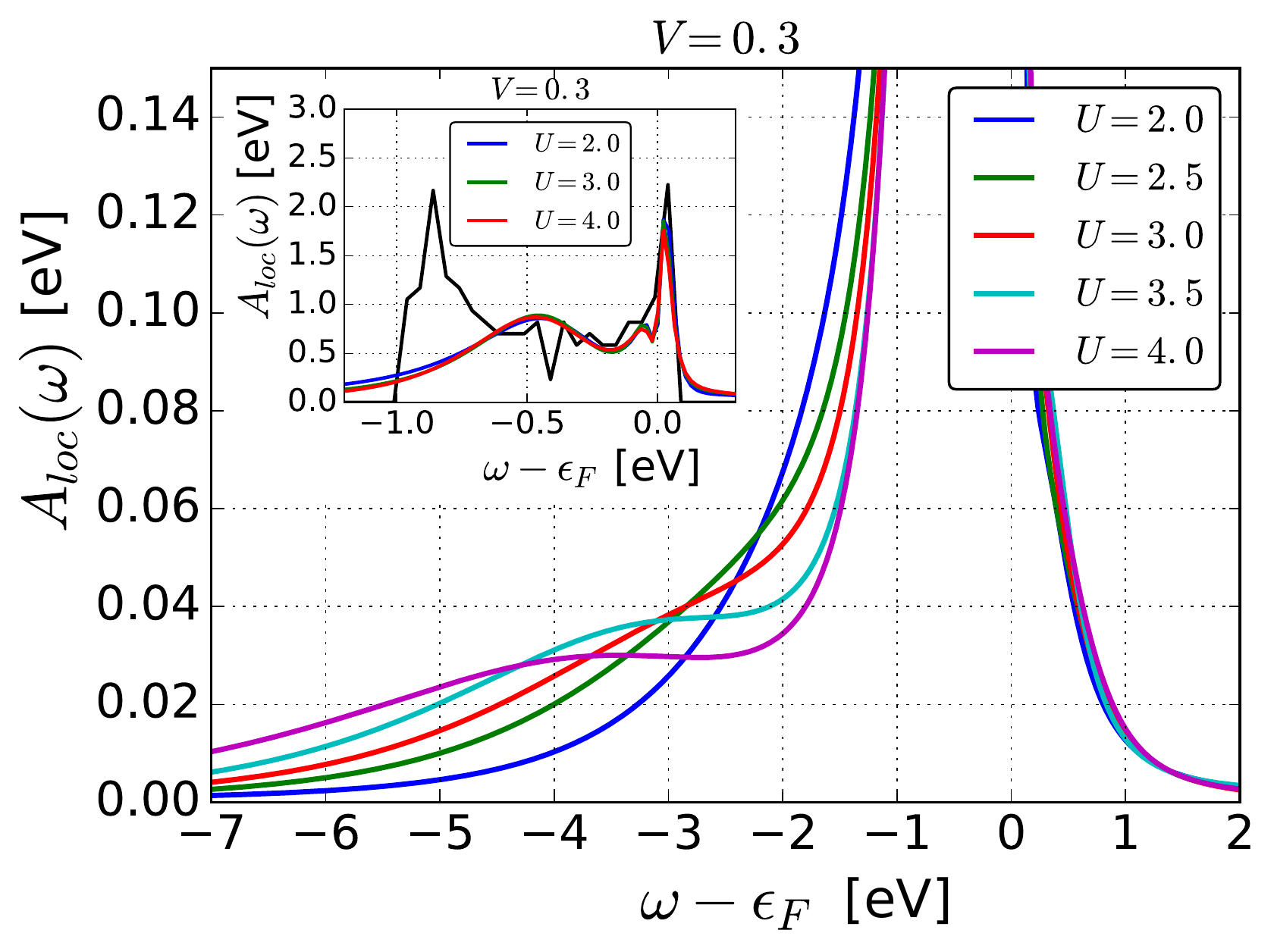}
\caption{Local spectral functions from EDMFT, calculated as $A_{\text{loc}}(\omega) = \sum_{\mathbf{k}} A(\mathbf{k}, \omega)$. The main and inset panels display the same functions on a different scale. Black: the non-interacting density of states, as in Fig.~\ref{hoppings}. Inset: The renormalization of the quasi-particle is visible on the total quasi-particle width. Main: A broad lower Hubbard band is visible, at energies between -4 eV and -3 eV. There is no visible upper Hubbard satellite.}
\label{spectra}
\end{center}
\end{figure}
\begin{figure}[htb]
\begin{center}
\includegraphics[width=1.0\columnwidth]
{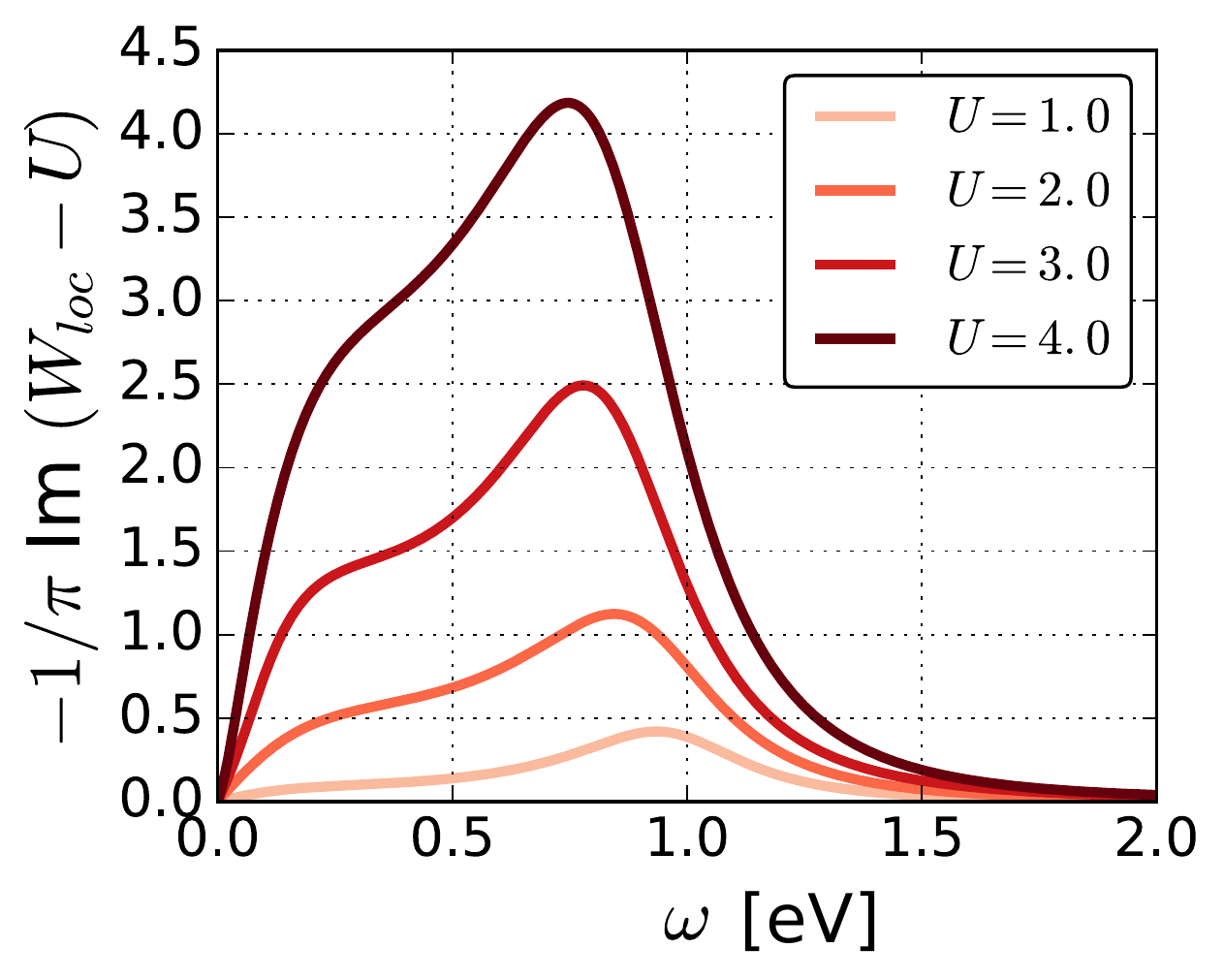}
\caption{Local part of the screened interaction, on the real
frequency axis, as calculated from an analytic continuation
using the Pad\'e scheme.}
\label{continuation_W_loc}
\end{center}
\end{figure}

Fig.~\ref{spectra} shows local spectral functions, in EDMFT, defined as $A_{\text{loc}}(\omega) = - \frac1{\pi} \text{Im} G_{\text{loc}}(\omega)$ and obtained via analytic continuations to the real-frequency axis using the maximum-entropy algorithm. In the inset, the spectra are represented on top of the non-interacting density of states. In the interacting spectrum, two quasi-particle structures -- one narrow and one broad -- are visible, corresponding to the two
van Hove singularities in the DOS. As can be expected from the values of $Z$ and since the self-energy is local, the broader quasi-particle feature is renormalized (resulting in a shift of its maximum towards the Fermi level) as compared to the non-interacting case, and
the bandwidth is reduced. On the main figure, the same spectra are zoomed in, to highlight a lower Hubbard band for high values of $U$. There is no upper Hubbard band. We interpret this asymmetry in the spectral function as a signature of the strong doping. Indeed, if one imagines a finite but large system, completely filled except for one hole, then the photoemission spectrum is going to be very asymmetric. On the one hand, hole-removal (or, equivalently, electron-addition) energies exactly correspond to the non-interacting ones, hence the absence of an upper Hubbard band. On the other hand, hole-addition (or, equivalently, electron-removal) energies have to take into account the interaction between two holes, hence creating a lower Hubbard band.
Note that one can distinguish this Hubbard band from a satellite that would originate from electron-hole excitations contained in the frequency-dependent screening as it is the case e.g. in GW, since the latter displays structure  below 1 eV (see Fig.\ref{continuation_W_loc}), whereas the hole-hole interaction is of the order of $U$, consistent with the observed distance from the quasi-particle feature (Fig.\ref{spectra}).

\section{Two-particle observables}
\label{section:two_part}

In this section, we describe our results for two-particle quantities corresponding to neutral charge excitations, in the homogeneous Fermi-liquid phase of the phase diagram Fig.~\ref{phase_diag_comp_fock}. We argue that the two-particle observables are typical of a dilute system, with few charge carriers. They appear uncorrelated, as they do not retain traces of the Hubbard satellite present in the single-particle quantities. We first describe the impurity susceptibility, screened and effective bare interactions, and then turn to a more detailed analysis of the particularities of the high-doping regime.

\begin{figure}[!htb]
\begin{center}
\includegraphics[width=1.0\columnwidth]{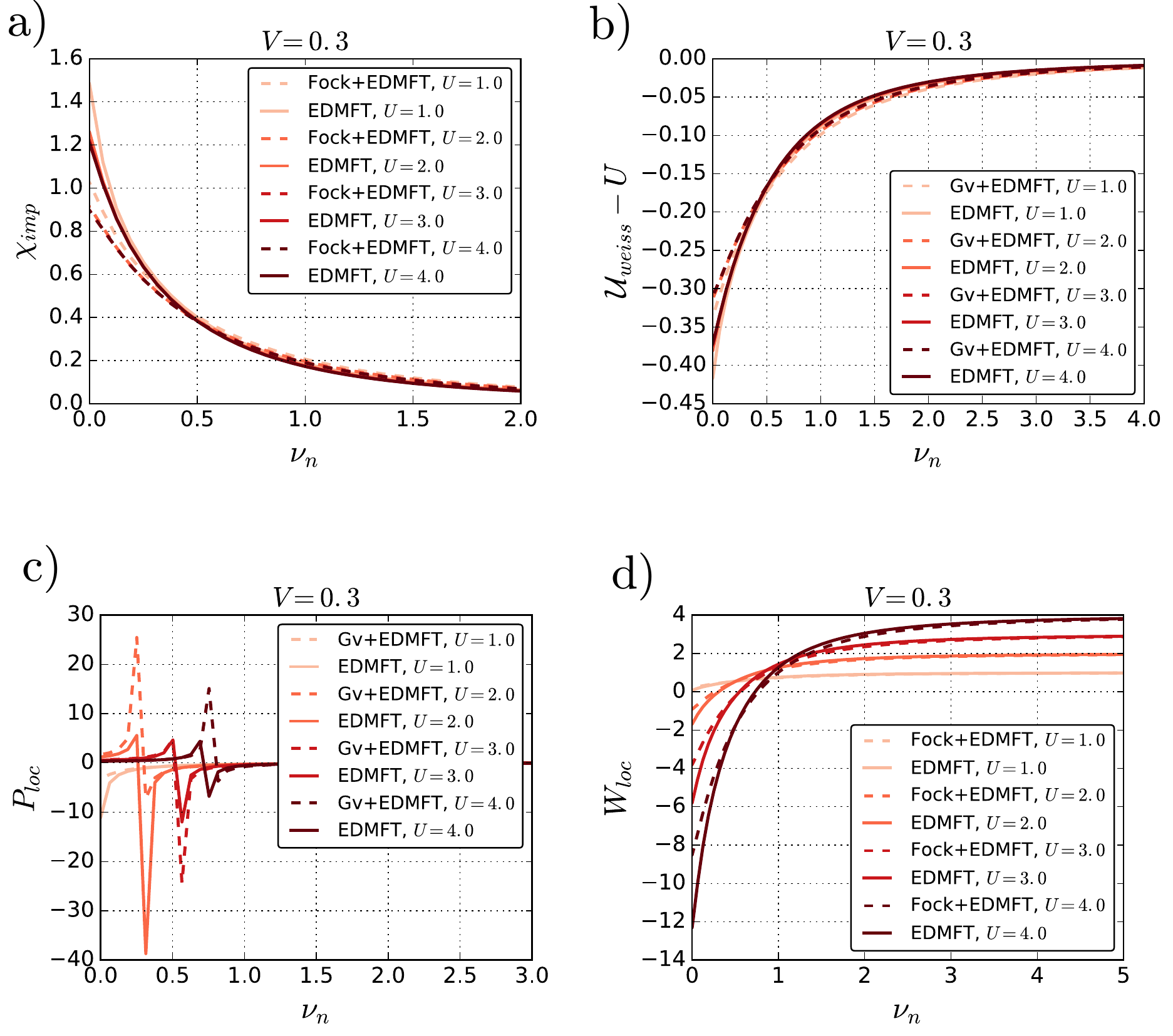}
\caption{Comparison between EDMFT and Fock+EDMFT for various two-particle quantities on the Matsubara axis. a) The impurity susceptibility, $\chi_{\text{imp}}(i\nu_n)$. For each method (EDMFT or Fock+EDMFT) the susceptibility is almost constant as a function of $U$. The susceptibilities within Fock+EDMFT are smaller than the susceptibilities within EDMFT. b) The dynamical interaction minus the local interaction, $\mathcal{U}(i\nu_n) - U$. c) The impurity polarization, $P_{\text{imp}}(i\nu_n)$. For the analyzed values of $U$ and $V$, the polarization displays a pole on the Matsubara axis. d) The local part of the screened interaction, $W_{\text{loc}}(i\nu_n)$. The $\nu_n=0$ value can become negative for large enough values of $U$.}
\label{two_part_fock}
\end{center}
\end{figure}

\begin{figure}[htb]
\begin{center}
\includegraphics[width=1.0\columnwidth]
{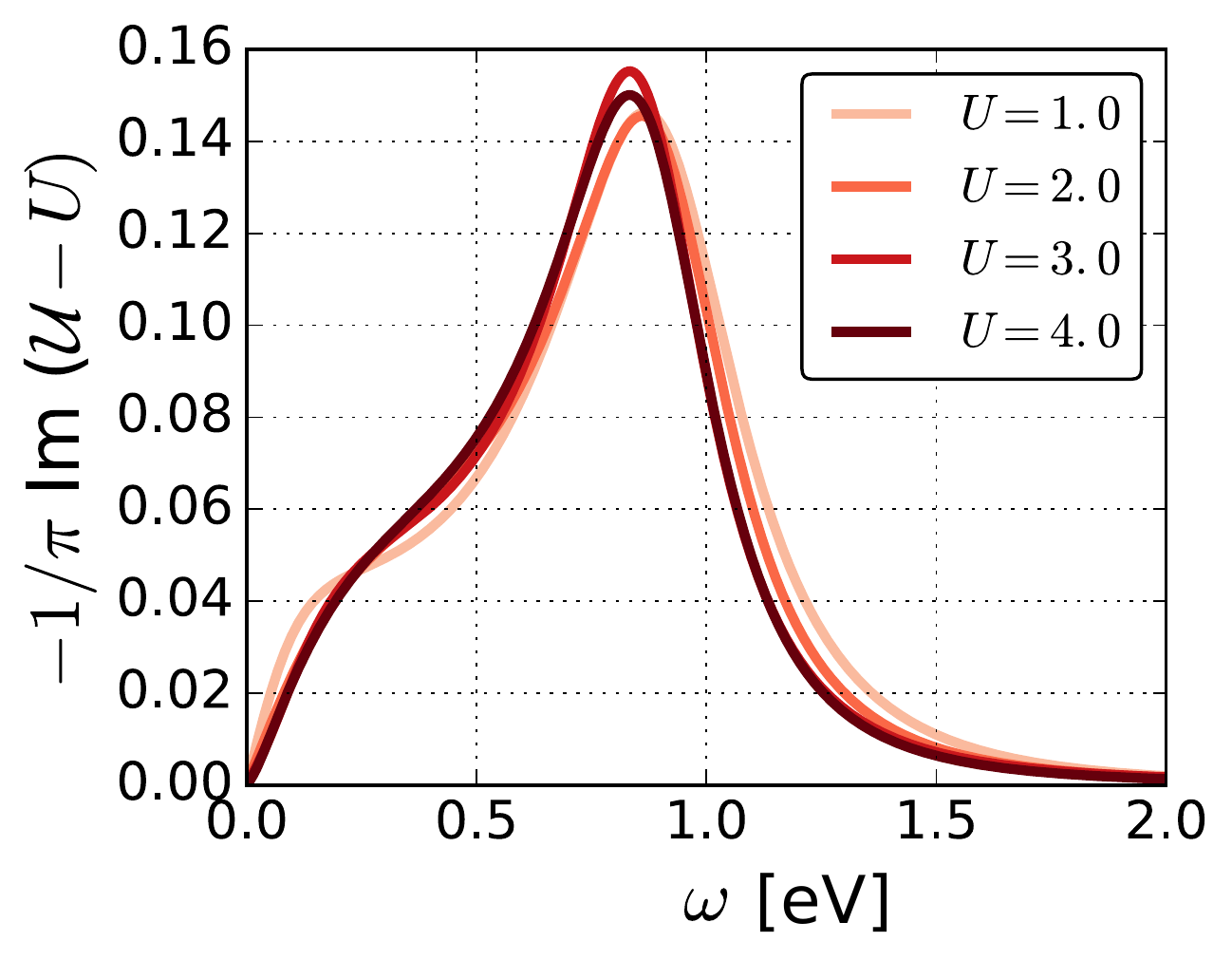}
\caption{Effective dynamical interaction 
on the real
frequency axis, as calculated from an analytic continuation
using the Pad\'e scheme.}
\label{continuation_U_weiss}
\end{center}
\end{figure}

\subsection{The impurity susceptibility}

The local charge susceptibility 
$\chi_{\mathrm{imp}}(\tau) = \braket{\mathcal{T} n(\tau)n(0)} - \braket{n}^2$
has been computed from the Anderson impurity model in imaginary time.
It is thus a single-site dynamical quantity, which is positive on the 
imaginary time and Matsubara frequency axes. Its Matsubara axis representation
is plotted in Fig.~\ref{two_part_fock}(a) for several values of $U$,
both within pure EDMFT and Fock+EDMFT. As a bosonic quantity, it decays
for large frequencies as $\frac{1}{\nu^2}$. System-specific information
is therefore rather contained in the low-frequency behavior.
The most striking features of the plot are 
(i) the weak dependence on $U$, indicating that the impurity charge 
susceptibility is only weakly renormalized by correlations and 
(ii) the marked difference between the EDMFT and Fock+EDMFT
results. We will analyze both of these points in detail below.

\subsection{Effective local interaction}

The frequency-dependent interaction, $\mathcal{U}(i\nu_n)$, is represented 
in Fig.~\ref{two_part_fock}(b). The real-frequency representation of its
imaginary part (obtained by analytic continuation using Pad\'e 
approximants) is plotted in Fig.~\ref{continuation_U_weiss}.
This interaction, a partially screened interaction, is an auxiliary
quantity of the EDMFT scheme. It effectively mimics the effect of the
non-local interaction $v_{\mathbf k}$ onto the local interaction, by 
introducing a frequency-dependence. The imaginary part can -- up to
a factor $- \pi$ -- be understood as the density of screening modes thus
generated.
The Matsubara axis plot Fig.~\ref{two_part_fock}(b) gives the difference
between $\mathcal{U}(i\nu_n)$ and the bare interaction parameter $U$ of
the model. Since at high enough frequencies screening is no longer
effective, $\mathcal{U}(i\nu_n)$ goes to $U$ in this limit, and the 
difference $\mathcal{U}(i\nu_n)-U$ vanishes.
At zero-frequency, $\mathcal{U}(i\nu_n = 0) = U_{\text{eff}} < U$ goes 
to a screened static value. 
At low frequencies, 
$\tilde{\mathcal{U}} (i\nu_n) = \mathcal{U}(i\nu_n)-U$ 
is a measure of the screening
induced by the non-local interactions. In general, the efficiency of
screening depends on the charge-charge correlations, which can be strongly
influenced also by the local part of the bare interaction, the Hubbard $U$.
Here, we find however that this partial screening depends only weakly 
on the local $U$: The curves in both, Fig.~\ref{two_part_fock}(b)
and Fig.~\ref{continuation_U_weiss}, show a quite negligible $U$-dependence:
To a good approximation, 
$\tilde{\mathcal{U}} = \tilde{\mathcal{U}}[V, \chi_0]$, i.e. it 
depends only on $V$ and on the band structure, but not on the local
interaction $U$.
Below, we will argue that this finding is a consequence of the weak
$U$-dependence of the impurity charge susceptibility found above.

\begin{figure}
\begin{center}
\includegraphics[width=1.0\columnwidth]{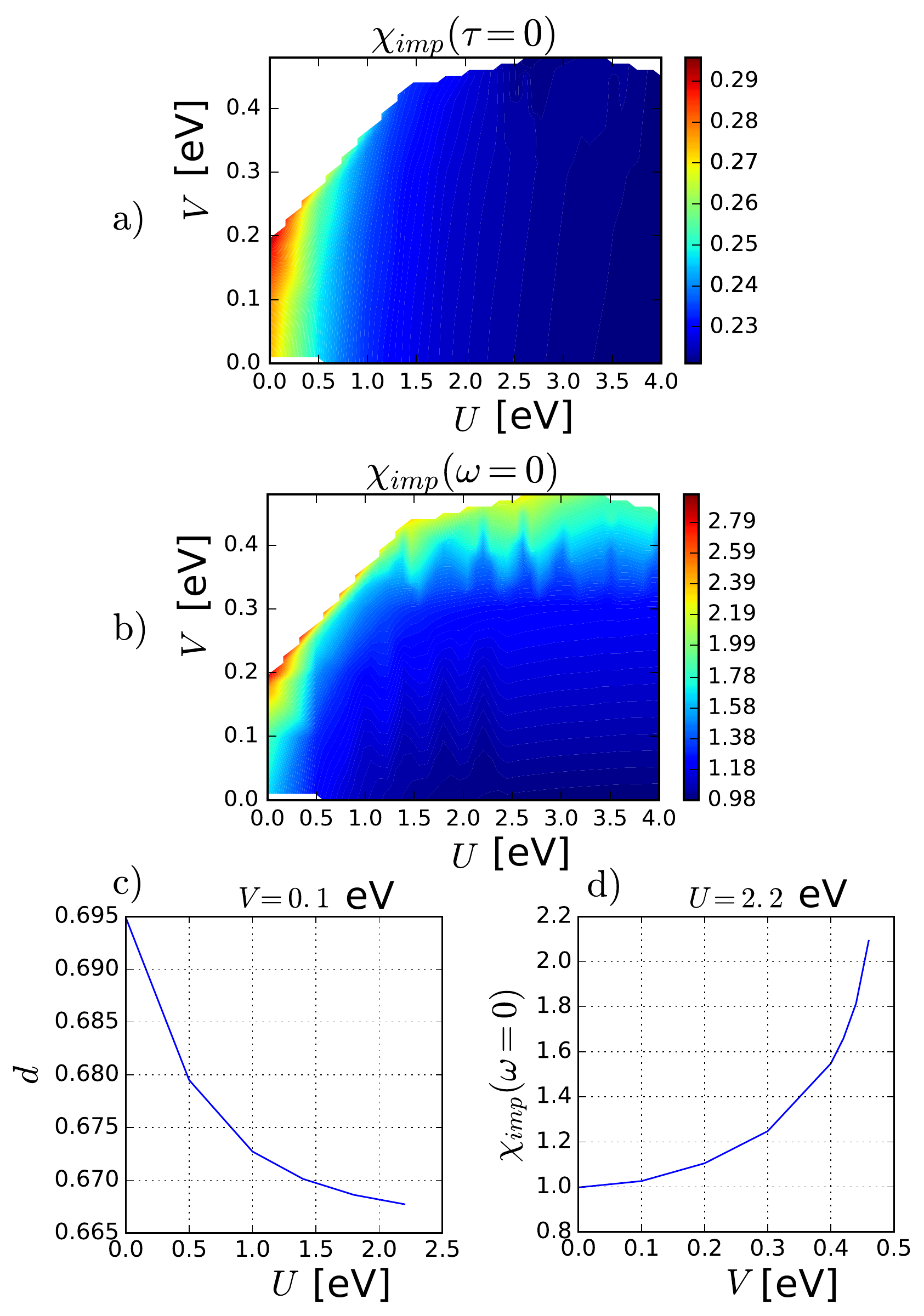}
\caption{Analysis of the impurity susceptibility, $\chi_{\text{imp}}$ in EDMFT. a) Colour plot of $\chi_{\text{imp}}(\tau=0)$ over the whole phase diagram. $\chi_{\text{imp}}(\tau=0)$ is directly connected to the double-occupancy, $d$. b) Colourplot of $\chi_{\text{imp}}(\omega=0)$ over the whole phase diagram. A divergence in $\chi_{\text{imp}}(\omega=0)$ signals a second-order instability. c) Double-occupancy $d$ as a function of $U$ for constant $V$. d) $\chi_{\text{imp}}(\omega=0)$ as a function of $V$ for constant $U$.}
\label{analyse_chi_edmft}
\end{center}
\end{figure}

\subsection{Screened Coulomb interaction and polarization}

Fig.~\ref{two_part_fock}(d) shows the fully screened Coulomb interaction
on the Matsubara axis, Fig.~\ref{continuation_W_loc} its imaginary
part on the real axis. The overall shape of these quantities is similar
to that of the effective interaction $\mathcal{U}$ discussed above, but the
comparison of the two quantities is instructive:  
$ W = \frac{\mathcal{U}}{1-P_{\text{imp}} \mathcal{U}}$ results from 
screening the effective $\mathcal{U}$ by the local impurity polarization
(see Fig.~\ref{two_part_fock}(c)): this additional screening leads to an overall
reduction of $W$ as compared to $\mathcal{U}$, which becomes obvious
already from the different scales. As expected, the energy range 
where screening modes exist stays roughly the same (compare 
Fig.~\ref{continuation_U_weiss} and Fig.~\ref{continuation_W_loc})
but the coupling strength of these modes is much enhanced. Also,
a more pronounced shoulder towards low frequencies in $W$ indicates
an enhancement of low-energy modes.
Such modes have also been found in the density-density response of the
homogeneous electron gas at low densities and large wave vectors
\cite{Takada}. Note that the local screening as included in EDMFT
corresponds to an average over all wave vectors.
Finally, the overall screening strength now does depend quite strongly
on $U$ (see below). 

\begin{figure}
\begin{center}
\includegraphics[width=0.85\columnwidth]{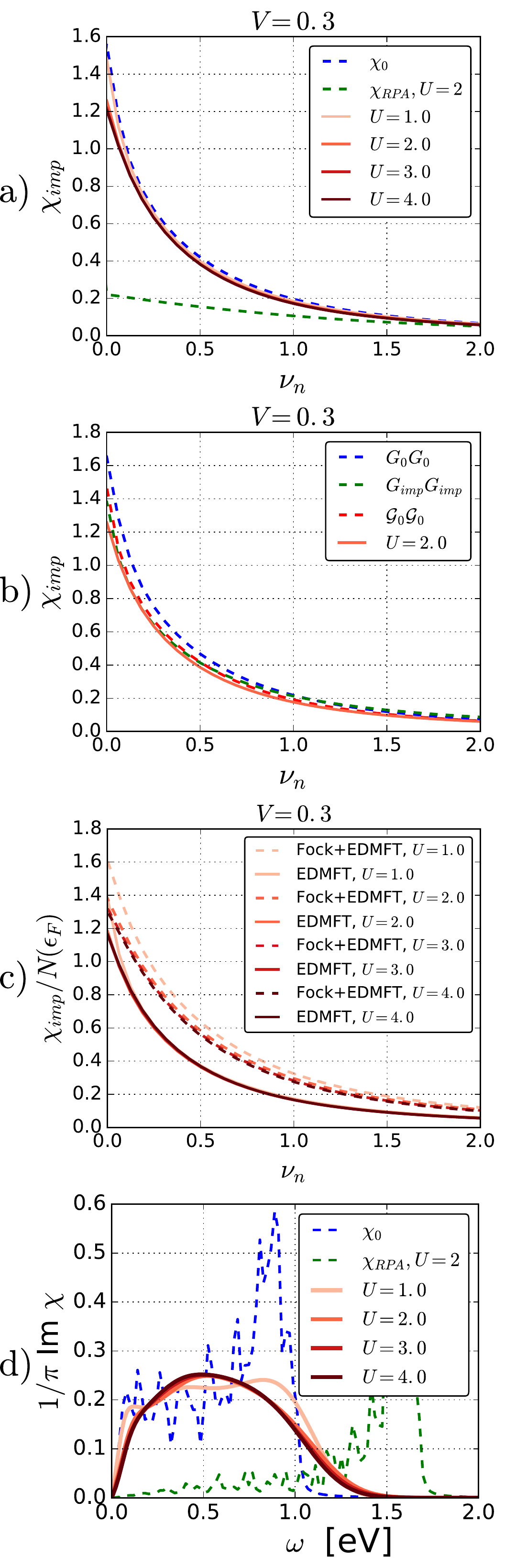}
\caption{
Analysis of the impurity susceptibility, $\chi_{\text{imp}}$. a) $\chi_{\text{imp}}(i\nu_n)$ on the Matsubara axis within EDMFT. It is close to its non-interacting counterpart, $\chi_0 = G_0 G_0$, even for large values of $U$. b) Comparison to various (local) bubble diagrams. c) Comparison between EDMFT and Fock+EDMFT susceptibilities, both renormalized by the effective interacting density of states at the Fermi level, $N(\epsilon_F)$ (see text). d) Analytic continuation of the EDMFT susceptibility, compared to the non-interacting susceptibility $\chi_0 = 2 G_0 G_0$, and the random-phase approximation (RPA) susceptibility, $\chi_{\text{RPA}}$.}
\label{divers_chi_edmft}
\end{center}
\end{figure}

\subsection{Analysis of the impurity susceptibility}

The weak renormalization by correlations can be understood by analyzing two components of the susceptibility: the frequency-integrated value $\chi_{\text{imp}}(\tau=0)$ and the static value $\chi_{\text{imp}}(\omega=0)$, both represented in Fig.~\ref{analyse_chi_edmft}, spanning various parameter sets throughout the phase diagram. $\chi(\tau=0)$ (the area under  the curve $\chi(i\omega)$) is linked to the double-occupancy on the impurity, $d$, via:
\begin{equation}
d = 
\frac12 \left( \chi_{\text{imp}}(\tau=0) + \braket{n_{\uparrow} + n_{\downarrow}}^2 - \braket{n_{\uparrow} + n_{\downarrow}} \right).
\end{equation}The double-occupancy goes from a non-interacting value (0.83*0.83) at low $U$ (the equivalent of 0.25 at half-filling), to an interacting value 1*(1.667-1) at high $U$ (the equivalent of 0 at half-filling). However, due to the filling constraint, these two values are numerically close. $\chi(\omega = 0)$ is also represented. It becomes maximum at the charge-ordering transition line, indicating a second-order instability.

The impurity susceptibility is thus weakly renormalized by the local interaction $U$. In fact, in EDMFT, the susceptibility is close to its non-interacting counterpart. Let us define the local non-interacting lattice susceptibility, $\chi^0_{ii}(\tau) = 2 G^0_{ii}(\tau) G^0_{ii}(\beta-\tau)$, where $G^0_{ii}$ is the local lattice non-interacting Green's function. A comparison between the impurity susceptibility and the local non-interacting lattice susceptibility is presented in Fig.~\ref{divers_chi_edmft}. These quantities are numerically close to each other. In Fig.~\ref{divers_chi_edmft}(b) comparison is also presented with a bubble composed of interacting Green's functions, $2 G_{\text{imp}} G_{\text{imp}}$, and at the impurity level with $2 \mathcal{G}_0 \mathcal{G}_0$. All these quantities are numerically close.

\subsection{Comparison of EDMFT and Fock+DMFT}

The effect of the Fock term on the susceptibilities can be seen in Fig.~\ref{two_part_fock}. The Fock+EDMFT susceptibilities are significantly smaller than the EDMFT ones. Indeed, as we have seen, the non-local Fock self-energy effectively widens the non-interacting band. This means, in particular, that $N(\epsilon_F)$ decreases, which leads to a decrease in $\chi$ (as a reminder, for a non-interacting metal, the static susceptibility goes as $N(\varepsilon_F)$). In Fig.~\ref{divers_chi_edmft}, $\chi/N(\epsilon_F)$ is represented, for both EDMFT and Fock+EDMFT. $N(\epsilon_F)$ is estimated from the formula $-\beta/\pi G_{\text{loc}}(\beta/2)$, as in Fig.~\ref{single_part_fock}. Hence, the renormalization of the susceptibility upon addition of the Fock term can be traced back to the decrease of the density of states at the Fermi level. This decrease of the impurity susceptibility implies that when the Fock term is added, the susceptibility differs more from the non-interacting one than in the case of pure EDMFT.

Finally, we show analytic continuations of the EDMFT impurity susceptibility. The spectra retain the shape expected from the non-interacting density of states. There is no sign of the Hubbard satellite, which was present in the single-particle spectra. 
Two aspects may contribute to this finding: first, in the framework of Hedin's equations,
satellites in two-body spectra can be rationalized as stemming from the frequency-dependence of both the
self-energy and of its derivative $\delta \Sigma/\delta G$, which contains the electron-hole interaction. From the framework
of calculations based on the Bethe-Salpeter equation \cite{PhysRevLett.78.1528} and from cumulant expansions of the electron-hole Green's 
function \cite{zhou-jcp2015}
it is known that these two contributions have a tendency to cancel each other. Second, 
in a dilute system, neutral excitations (with fixed particle number) correspond to non-interacting excitations. Thus, 
cancellations must be particularly efficient
The weak deviation of $\chi_{\text{imp}}$ from its non-interacting counterpart can thus be 
linked to the high doping level, and one may expect
a qualitatively similar behaviour for the lattice charge dynamics.

\subsection{Consequences for the lattice charge susceptibility}

As discussed in the methods section, when adopting a ``purist's point of
view'', EDMFT does not give access to the lattice susceptibility.
Nevertheless, keeping the limitations of the local approximation in
mind, one may still ask the question of how this quantity would look
like when calculated under the assumption of a purely local polarization
function given by the EDMFT one, see Eq.~(\ref{susceptibility_edmft}).

Inspired by the analysis of the impurity polarization and the effective
interaction above, we make the simplifying assumption that 
$\chi_{\text{imp}}$ is independent of $U$ or $V$ through the whole phase 
diagram and that $\chi_{\text{imp}} = \chi^0$. As is clear from the above
analyses, the qualitative behaviour of all auxiliary quantities can be 
well understood in terms of this simplifying assumption.
Here, we analyze 
the implications for the (Fock +) EDMFT lattice susceptibility. 
Let us decompose $v_{\mathbf k} = U + \Delta v_{\mathbf k}$, such that $\Delta v_{\mathbf k}$ depends only on the intersite interaction $V$. Assuming that $\chi_{\text{imp}} = \chi_0$, the EDMFT susceptibility, Eq.~\eqref{susceptibility_edmft}, can be re-expressed as:
\begin{equation}
\chi^{\text{EDMFT}} (\mathbf{k}, \omega) =
\frac{\chi_0}{1 - \tilde{\mathcal{U}} \chi_0 + \Delta v_{\mathbf k} \chi_0}.
\end{equation}
Since $\tilde{U} = \mathcal{U} -U$ is almost independent of $U$ 
(see Fig.~\ref{two_part_fock}), the same holds for
the EDMFT susceptibility.

\subsection{Negative screened interaction}

Finally, we report a region characterized by a negative static screening. As already mentioned in section~\ref{sec:phase_diag}, in the phase diagram, Fig.~\ref{phase_diag}, the Fermi-liquid phase is divided into a region where the static local screening is positive, $W_{\mathrm{loc}}(\omega=0) >0$ at small $U$'s, and a region where the static local screening is negative, $W_{\mathrm{loc}}(\omega=0) <0$ at large $U$'s. The qualitative frequency-dependence of $P_{\text{imp}}$ and $W_{\text{imp}}$ is a direct consequence of the approximate equality $\chi=\chi_0$. Indeed, assuming that $\chi_{\mathrm{imp}}(\omega) = \chi_0(\omega)$, $P_{\mathrm{loc}}$ and $W_{\mathrm{loc}}$ can be re-expressed as:
\begin{align}
&P_{\mathrm{loc}} = - \frac{\chi_0}{1 - \mathcal{U} \chi_0} \label{eq:Ploc}\\
&W_{\mathrm{loc}} = \sum_{\mathbf k} \frac{v_{\mathbf k}}{1 - v_{\mathbf k} P_{\mathrm{imp}}} 
= (1-\mathcal{U} \chi_0) \sum_{\mathbf k} \frac{v_{\mathbf k}}{1 + v_{\mathbf k} \chi_0}.
\label{eq:Wloc}
\end{align}
Eqs.~\eqref{eq:Ploc}~and~\eqref{eq:Wloc} show that, depending on the value of $1-\mathcal{U}\chi_0$, there is a possibility for both a pole in $P_{\text{loc}}$ and a sign-change in $W_{\text{loc}}$. These quantities are represented in Fig.~\ref{two_part_fock}. For $V=0.3$ and $U=2$, 3, 4, $P_{\text{loc}}$ displays a pole at finite frequency, while $W_{\text{loc}}$ changes sign. In particular, at zero-frequency, $W_{\text{loc}}$ takes a negative value.

The possibility that the local static screening is negative is not
an artefact of the (Fock+) EDMFT approximation. It is a feature of any
system where the susceptibility is large compared to the inverse interaction:
$|\chi| v_{\bf k}>1$. 
In appendix A, we show that such a situation is easy to generate by
presenting an example of a simple exactly solvable model where
$W_{\mathrm{loc}}$ becomes negative.
In the EDMFT context, negative screened interactions have also been observed 
in the context of an extended
Hubbard model on the square lattice \cite{Huang2014}.
If -- with the caveats above -- one interprets the EDMFT dielectric
function as the physical dielectric function of the system, we notice
that a negative {\it local} static $W$ requires a set of $q$-vectors for
which the inverse dielectric function $1/\epsilon(q,\omega=0)$ is negative.
In the present case, due to the structure of the bare interaction,
the negative values extend over the whole first Brillouin zone, including
$q \rightarrow 0$, corresponding to
a negative electronic compressibility of the system.

Further situations where negative static inverse dielectric functions
might appear have been discussed in the literature: a prominent example
is the jellium model, an electron gas with a static uniform positive
compensating background \cite{Nozieres}, in the low density regime.
However, the interpretation of this effect remains subtle.
Nozi\`eres \cite{Nozieres} argues that in the presence
of a negative electronic compressibility, the system would become
unstable with respect to density fluctuations of the positive background.
This would suggest that in the solid, with potentially mobile ionic
degrees of freedom, the relevant quantity to analyze is the total
(electronic plus ionic) compressibility rather than the electronic
one alone. Explicit calculations for simple metals have been performed
by Kukkonen and Wilkins \cite{kukkonen}, where inclusion of polarization
effects of the ionic cores was shown to be important.
Dolgov et al. \cite{RevModPhys.53.81}, on the other hand, stress that a negative
dielectric function {\it at finite $q$-vectors} does not {\it a priori}
contradict the requirements of system stability. These authors review
a variety of different physical systems where such a situation
appears.
The electron gas problem was taken up in detail by Takada \cite{Takada}
and Takayanagi and Lipparini \cite{Takayanagi}, who have investigated the
finite $q$-behavior resulting from the negative dielectric function of
the dilute electron gas: in this regime, a collective mode (``ghost
plasmon'' or ``ghost exciton'') was identified.

In the present case, we are dealing with a region of negative
$\epsilon(q,\omega=0)$ extending over the whole first Brillouin zone,
thus formally corresponding to the case of negative electronic
compressibility. One may speculate that the present regime indeed already
corresponds to a situation where the positive ionic lattice would become
unstable towards lattice distortions or phase separation.

To conclude this section, we have shown that both quantities, $P_{\text{loc}}$ and $W_{\text{loc}}$, self-consistently adjust in such a way that the approximate equality $\chi_{\text{imp}} = \chi_0$ is preserved. This may create a pole in the impurity polarization on the Matsubara axis. Importantly, this may cause the value of the static screening to become negative and effectively attractive.

\section{Summary and Conclusions}

We have studied the extended Hubbard model on the triangular lattice
in a regime of high doping, close to the band-insulating limit. 
We have briefly described the EDMFT and the
Fock+EDMFT schemes. In the latter, a non-local Fock self-energy diagram
supplements the purely local diagrams of EDMFT. We have computed the
phase diagram as a function of local and non-local interactions. For low 
intersite interactions, a homogeneous metallic phase is found. 
For high intersite interactions, the homogeneous phase could not be
stabilized, hinting at a charge-ordered, symmetry-broken phase. 
In the homogeneous metallic phase, a region with negative local static
screening is observed. The negativity of the static screened interaction can
be traced back to the high doping, the large on-site interaction
and the band structure.
Within the EDMFT approximation, the present situation of a local
negative screened interaction goes hand in hand with a negative
compressibility (see Section V.G),
signaling a possible instability of the positive background.
Its consequences are an interesting open question, that would however
require an extension of the model in order to include ionic degrees
of freedom explicitly.

We have analyzed spectral properties within (Fock +) EDMFT in the
homogeneous metallic phase at the level of one- and two-particle observables, 
rationalizing
seemingly contradicting trends. The one-body spectral function is asymmetric
owing to the high doping, and features a lower Hubbard band as a result of correlations. On the two-body level,
the charge susceptibility is found to be weakly renormalized by interactions. This observation,
using a non-perturbative methodology, demonstrates a remarkable failure of the
random-phase approximation in our particular doping regime. Indeed, the interacting
susceptibility is even less renormalized by correlations than within the RPA.

\section{Perspectives}

Understanding the effects of non-local interactions is of course
also an issue that is relevant from a materials science perspective.
Charge-ordering is an ubiquitous phenomenon in (quasi-)
two-dimensional materials with a triangular lattice geometry:
examples include 
sodium and lithium cobaltates~\cite{fouassier_JSSC_73, amatucci_JTES_96}, 
two-dimensional organics~\cite{hotta_C_12, kagawa_NP_13} or 
surface systems~\cite{Hansmann2013, Hansmann2016}. 

The non-interacting part of our model parameters was chosen to be 
representative of a specific system, namely Na-doped cobaltates. 
These two-dimensional systems display a rich phase diagram
as a function of doping~\cite{foo_PRL_04, lang_PRB_08, schulze_PRB_08}. 
Interestingly, many of the intriguing properties that can be
assigned to correlations occur at high doping, near the band-insulating 
regime. In particular, in the high-doping regime, it was found that the
triangular lattice disproportionates into a lattice with a four-fold
larger unit cell, forming a Kagome lattice of active sites \cite{alloul}.
A charge ordering pattern corresponding to this symmetry breaking
would correspond to an ordering vector $\Gamma-M$.
Considering any realistic set of parameters $U$ and $V$ for the 
cobaltates (typically $U \geq 2$ eV~\cite{hasan_PRL_04}), within EDMFT, the
system falls either in the negative-screening or the charge-ordered region.

In the future, it would therefore be interesting to extend
our study by allowing symmetry-broken solutions in real space. 
Indeed, a still open interesting question is the origin of the 
experimentally observed charge ordering in the cobaltates,
including its ordering vector. The wave vector suggested by the
present EDMFT calculations does not allow to understand the experimental
situation.

From the methodological point of view, it could be interesting to benchmark our
phase diagram against extensions of EDMFT. In particular, the contribution from
non-local self-energy and polarization diagrams could be checked by comparing 
our results to 
$GW$+DMFT~\cite{Biermann2003,Sun2004,Ayral2012,Ayral2013,Biermann2014}, 
TRILEX~\cite{ayral_PRB_15, ayral_PRB_16,Ayral2017b,Vucicevic2017},
dual boson \cite{VanLoon2014} or cluster calculations \cite{PhysRevB.95.115149}.

\section*{Acknowledgments}

We thank L. Boehnke, E. van Loon and M. Panholzer for useful discussions.
This work was supported by the European Research Council 
(Projects CorrelMat, grant agreement 617196 and SEED, grant agreement 320971)
and IDRIS/GENCI Orsay (Project No. t2017091393).

\section*{Appendix: Negative screening in the Hubbard dimer}
\label{anx:dimer}
In this appendix, we further investigate the question of the
appearance of a negative local screened Coulomb interaction.
In particular, we show that a negative static screening is not 
an artefact of the EDMFT approximation, but
is expected in this low-density regime.

We show in the following that it appears in the
exact solution of specific many-body systems.

Let us define the Hubbard dimer at 1/4-filling via the Hamiltonian:
\begin{equation}
H = U \sum_{i=1,2} n_{i\uparrow} n_{i\downarrow}
- t \sum_{\sigma} (c_{1\sigma}^{\dagger} c_{2\sigma} + c_{2\sigma}^{\dagger} c_{1\sigma})
- \mu \sum_i n_i,
\end{equation}
where $\mu$ is chosen such that the many-body ground state features one electron.

The charge susceptibility can be computed exactly and does not depend on the interaction $U$ at quarter-filling. On the Matsubara axis, the local (diagonal) component is:
\begin{align}
\chi_{11}(i\nu_n) &= -\frac14 \left[ \frac1{i\nu_n - 2t} - \frac1{i\nu_n +2t} \right] \nonumber\\
&= \frac{t}{\nu_n^2 + 4t^2}.
\end{align}
This is a lorentzian centered around 0. The value at the Matsubara frequency $\nu_n =0$ is:
\begin{equation}
\chi_{11} (i\nu_n = 0) = \frac1{4t},
\end{equation}
which goes to infinity when $t \rightarrow 0$. With the formula $W_{ij} = v_{ij} - v_{ik} \chi_{kl} v_{lj}$, where $i, j, k, l$ denote site indices, we find that for $\nu_n = \omega = 0$:
\begin{equation}
W_{11}(\omega=0) = W_{\text{loc}}(\omega=0) = U - \frac{U^2}{4t},
\end{equation}
which becomes negative as soon as $U/(4t) >1$. Hence, the fact that the local part of $W$ becomes negative is not an artifact of an approximation. It happens in our exactly-solvable model.

The off-diagonal (intersite) element of the susceptibility is given by: $\chi_{12} = - \chi_{11}$ (as it should, given the requirement of charge conservation, which imposes $\chi(q=0,0)=0$). From this, the local part of the polarization can be computed:
\begin{equation}
P_{11} = -\frac{\chi_{11}}{1-2U\chi_{11}} .
\end{equation}
In particular, when $\chi_{11}$ grows large, then $P_{11}$ develops a pole.

\bibliographystyle{apsrev4-1}

\end{document}